\documentclass[aps,floats]{revtex4}
\usepackage{amsmath,amssymb}
\usepackage{graphicx,epsfig}
\begin{document}
\bibliographystyle {plain}

\def\oppropto{\mathop{\propto}} 
\def\opsimeq{\mathop{\simeq}}
\def\opoverderline{\mathop{\overline}}
\def\operarrow{\mathop{\longrightarrow}}
\def\opsim{\mathop{\sim}}

\def\fig#1#2{\includegraphics[height=#1]{#2}}
\def\figx#1#2{\includegraphics[width=#1]{#2}}

%\newcommand{\fig}[2]{\epsfxsize=#1\epsfbox{#2}} \reversemarginpar 

%%%%%%%%%%%%%%%%%%%%%%%%%%%%%%%%%%%%%%%%%%%%%%%%%%%%%%%%%%%%%%%%%%%%%%%%%%%%
\title{Critical points of quadratic renormalizations of random variables \\
and phase transitions of 
disordered polymer models on diamond lattices } 

%%%%%%%%%%%%%%%%%%%%%%%%%%%%%%%%%%%%%%%%%%%%%%%%%%%%%%%%%%%%%%%%%%%%%%%%%%%%

 \author{ C\'ecile Monthus and Thomas Garel }
  \affiliation{Service de Physique Th\'{e}orique, CEA/DSM/SPhT\\
 Unit\'e de recherche associ\'ee au CNRS\\
 91191 Gif-sur-Yvette cedex, France}

\begin{abstract}
We study the wetting transition and the directed 
polymer delocalization transition on diamond hierarchical lattices.
These two phase transitions with frozen disorder 
correspond to the critical points
of quadratic renormalizations of the partition function.
( These exact renormalizations on diamond lattices can also be considered
as approximate Migdal-Kadanoff renormalizations for hypercubic lattices). 
In terms of the rescaled partition function $z=Z/Z_{typ}$,
we find that the critical point corresponds 
to a fixed point distribution with a power-law tail
 $P_c(z) \sim \Phi(\ln z)/z^{1+\mu}$
 as $z \to +\infty$ ( up to some sub-leading logarithmic
correction $\Phi(\ln z)$), so that all moments $z^{n}$ with $n>\mu$ diverge.
For the wetting transition, the first moment diverges $\overline{z}=+\infty$ 
(case $0<\mu<1$), and the critical temperature is strictly below
the annealed temperature $T_c<T_{ann}$.
For the directed polymer case, the second moment 
diverges $\overline{z^2}=+\infty$
(case $1<\mu<2$), and the critical temperature is strictly below
the exactly known transition temperature $T_2$ of the second moment.
We then consider the correlation length exponent $\nu$ :
the linearized renormalization around the fixed point distribution
coincides with the transfer matrix 
describing a directed polymer on the Cayley tree,
but the random weights determined by the fixed point
distribution $P_c(z)$ are broadly distributed. This induces some changes
in the travelling wave solutions
with respect to the usual case of more narrow distributions.

\end{abstract}

\maketitle

\section{ Introduction}

\subsection{Real-space renormalizations for disordered systems}

\begin{figure}[htbp]
%\begin{figure}
\includegraphics[height=6cm]{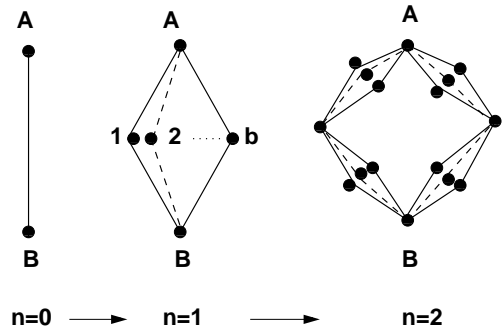}
\hspace{1cm}
\caption{ Hierarchical construction of the diamond lattice of
branching ratio $b$.   }
\label{figdiamond}
\end{figure}

The choice to work in real-space to define renormalization procedures, 
which already present a great interest for pure systems
 \cite{realspaceRG},
becomes the unique choice for 
 disordered systems if one wishes to describe spatial heterogeneities.
Whenever these disorder heterogeneities play a dominant role 
over thermal or quantum fluctuations, the most appropriate renormalizations
are strong disorder renormalizations \cite{StrongRGreview}
introduced by Ma-Dasgupta \cite{Ma-Dasgupta} : as shown by
Fisher \cite{daniel}, these strong disorder renormalization rules
lead to asymptotic exact results if the broadness of the disorder
distribution grows indefinitely at large scales.
However, for disordered systems governed by finite-disorder
fixed points, where disorder fluctuations remain of the same order
of thermal fluctuations, one needs to use more standard
real-space renormalization procedures, such as 
Migdal-Kadanoff block renormalizations \cite{MKRG}. 
  They can be considered in two ways, 
 either as approximate renormalization procedures on hypercubic lattices,
or as exact renormalization procedures on certain hierarchical lattices
\cite{berker,hierarchical}.
One of the most studied hierarchical lattice is the
diamond lattice which is constructed recursively
from a single link called here generation $n=0$ (see Figure \ref{figdiamond}) :
 generation $n=1$ consists of $b$ branches, each branch
 containing $2$ bonds in series ;
 generation $n=2$ is obtained by applying the same transformation
to each bond of the generation $n=1$.
At generation $n$, the length $L_n$ between the two extreme sites
$A$ and $B$ is $L_n=2^n$, 
the total number $B_n$ of bonds is $B_n=(2 b)^n=L_n^{d_{eff}}$
so that $d_{eff}(b)= \frac{ \ln (2b)}{\ln 2}$ represents
some effective dimensionality.

On this diamond lattice, various
disordered spin models have been studied,
such as for instance the diluted Ising model \cite{diluted}, 
random bond Potts model \cite{potts},
and spin-glasses \cite{hierarchicalspinglass}.
Disordered polymer models have also been considered,
in particular the  wetting on a disordered substrate
 \cite{Der_wett,Tang_Chate}
and the directed polymer model
 \cite{Der_Gri,Coo_Der,Tim,roux,kardar,cao, tang,Muk_Bha,Bou_Sil}.
In this paper, we focus on these two polymer models
that are described by quadratic renormalization
 of their partition functions as we now recall.

\subsection{Wetting transition with disorder on the diamond lattice}

On the diamond lattice, the adsorption of a polymer on a disordered substrate
is described by the following quadratic recursion
for the partition
function $Z_n$ of generation $n$  \cite{Der_wett}
\begin{eqnarray}
Z_{n+1} = Z_n^{(1)} Z_n^{(2)} + (b-1) Y_n^2
\label{zfullwetting}
\end{eqnarray}
where $Y_n= b^{L_n-1}$ represents the number
of walks between the two extreme points
and satisfies the recursion without disorder
\begin{eqnarray}
Y_{n+1}=b Y_n^2
\label{ynwetting}
\end{eqnarray}
and where $Z_n^{(1)}$ and $Z_n^{(2)}$ represent
two independent copies of generation $n$.
At generation $n=0$, 
the lattice reduces to a single bond
with a random energy $\epsilon$, for instance 
drawn from the Gaussian
distribution 
\begin{eqnarray}
\rho (\epsilon) = \frac{1}{\sqrt{2\pi} } e^{- \frac{\epsilon^2}{2} }
\label{gaussian}
\end{eqnarray}
and thus the initial condition for the recursion of
 Eq. \ref{zfullwetting}
is simply
\begin{eqnarray}
Z_{n=0} = e^{- \beta \epsilon}
\label{zrecursioninitial}
\end{eqnarray}
The temperature only appears in this initial condition.

\subsection{Directed polymer on the diamond lattice} 

The model of a directed polymer in a random medium \cite{Hal_Zha}
can also be studied on the diamond hierarchical lattice
 \cite{Der_Gri,Coo_Der,Tim,roux,kardar,cao, tang,Muk_Bha,Bou_Sil}.
The partition function
$Z_n$ of the $n$-generation satisfies the exact recursion  \cite{Coo_Der}
\begin{eqnarray}
Z_{n+1} = \sum_{a=1}^b Z_n^{(2a-1)} Z_n^{(2a)}
\label{zrecursion}
\end{eqnarray}
where $(Z_n^{(1)},...,Z_n^{(2b)})$ are $(2b)$ 
independent partition functions
of generation $n$.
At generation $n=0$, the lattice reduces to a single bond
with a random energy $\epsilon$, for instance 
drawn from the Gaussian of Eq. \ref{gaussian}
and thus the initial condition for the recursion of Eq. \ref{zrecursion}
is again given by Eq. \ref{zrecursioninitial}.

\subsection{ Organization of the paper }

In this paper, we study the critical points
of the quadratic renormalizations described above
that correspond to delocalization transitions for the polymer.
These two transitions
are of course different in nature, since the wetting transition
already exists in the pure case, whereas the directed polymer
transition only exists in the presence of disorder.
However, we will show below that the quadratic form of the renormalizations
induce some common properties. It is thus interesting to
study them along the same lines to stress their
similarities and differences.
The paper is organized as follows.
The wetting transition on a disordered substrate is discussed in Section
\ref{secwetting}, and studied numerically in Section \ref{secwettingnume}.
The directed polymer transition is discussed in Section
\ref{secdp}, and studied numerically in Section \ref{secdpnume}.
In Section \ref{comparison}, we compare the results on the diamond lattice with
respect to the same disordered polymer models defined on hypercubic lattices.
Section \ref{conclusion} contains the conclusion.
The Appendix \ref{multiplicative} contains a reminder on multiplicative
stochastic processes which is used in 
Sections \ref{secwetting} and \ref{secdp}.

\section{ Wetting on a disordered substrate }

\label{secwetting}

To study the wetting recursion,
it is convenient to introduce the reduced
partition function $z_n$ and the associated free-energy $f_n$
defined by \cite{Der_wett}
\begin{eqnarray}
z_{n} \equiv \frac{ Z_n} {Y_n} \equiv e^{- \beta f_n } 
\label{defzwetting}
\end{eqnarray}
to rewrite the recursion of Eq. \ref{zfullwetting} as
\begin{eqnarray}
z_{n+1} = \frac{z_n^{(1)} z_n^{(2)} + b-1}{b}
\label{zrecursionwetting}
\end{eqnarray}

\subsection{ Reminder on the pure case} 

\label{purewetting}

In the pure case, the ratios $z_n$ defined in Eq. \ref {defzwetting}
are not random but take a single value $R_n$, and the recursion of Eq. 
\ref{zrecursionwetting} reduces to a one-dimensional mapping $T$
\begin{eqnarray}
R_{n+1} = \frac{R_n^2 + b-1}{b} \equiv T(R_n) 
\label{mappingpurewetting}
\end{eqnarray}
discussed in \cite{Der_wett} : for $b>2$, there exists two attractive fixed
points $R_{\infty}=1$ (delocalized phase) and $R_{\infty}=+\infty$
(localized phase) separated by the repulsive fixed point $R_c$
(critical point) with
\begin{eqnarray}
R_c = b-1 
\label{rcpure}
\end{eqnarray}

The critical exponents are determined by the linearization of the 
recurrence around the fixed point. Setting $R_n=R_c + \delta_n$,
one obtains at linear order
\begin{eqnarray}
\delta_{n+1} \simeq \lambda \delta_{n} \ \ {\rm  with \ \ }
\lambda = T'(R_c) = \frac{ 2 R_c}{b} = \frac{ 2 (b-1)}{b}
\label{lambdapurewetting}
\end{eqnarray}

Note that this factor $\lambda=T'(R_c)>1$ 
describing the instability of the critical point
also governs the growing of the energy $E_n$ exactly at criticality
\cite{Muk_Bha}, since
the recursion for the energy
\begin{eqnarray}
E_{n+1} = \frac{ R_n^2 (2 E_n)}{R_n^2 + b-1 }
\label{epurewetting}
\end{eqnarray}
 becomes at criticality
\begin{eqnarray}
E_{n+1}(T_c) = \frac{ 2 R_c^2 }{R_c^2 + b-1 } E_n(T_c) = \lambda E_n(T_c) 
\label{ecritipurewetting}
\end{eqnarray}
To understand why the same factor $\lambda$ appears, 
one may introduce the product $U_n=R_n E_n$ 
that satisfies the recursion
\begin{eqnarray}
U_{n+1} = \frac{ 2 R_n }{b } U_n
\label{upurewetting}
\end{eqnarray}
It is then clear that at criticality it coincides with
 the recursion of the variables $\delta_n$ (Eq. \ref{lambdapurewetting}).

In conclusion, the variable $\delta_n$ or
 the energy $E_n$ at criticality grows as $\lambda^n = L_n^{1/\nu}$
in terms of the length $L_n=2^n$ with the critical exponent
\begin{eqnarray}
\nu = \frac{\ln 2}{ \ln \lambda} = \frac{\ln 2}{ \ln T'(R_c)}
\end{eqnarray}
The specific heat exponent satisfies the hyperscaling relation $\alpha=2-\nu$.
 We refer to \cite{Der_wett} for more details.

Let us now summarize the changes that the
presence of frozen disorder will induce :

(i) the one-dimensional mapping of the pure case $R_{n+1}=T(R_n)$ 
will become
 the iteration of a probability distribution $Q_{n+1}(z) = {\cal F } \{
Q_n(z) \}$

(ii) the critical value $R_c$ of the pure case will become an invariant 
probability distribution $Q_c(z)= {\cal F } \{ Q_c(z) \} $

(iii) the critical exponent $\nu$ determined by the derivative $T'(R_c)$
in the pure case will be determined by the linearized iteration
around the fixed point distribution $Q_c(z)$.

But before concentrating on the critical point, we first describe 
the properties of the renormalization group (RG) flow with disorder 
in the limits of high and low temperatures.

\subsection{ High-temperature RG flow} 

In the high temperature phase, the variables $z_n$ defined in Eq.
\ref{defzwetting} flow towards $1$ or equivalently
the free-energies $f_n$ decay to zero.
The linearization of the recursion in this regime yields
\begin{eqnarray}
f_{n+1} \opsimeq \frac{f_n^{(1)} +f_n^{(2)} }{b}
\label{recursionwettinghigh}
\end{eqnarray}
For $b>2$, this high-temperature phase exists,
the probability distribution of the free-energy converges to a
Gaussian, the average and the width decays as
power-laws of the length $L_n=2^n$
\begin{eqnarray}
\overline{ f_n} && \propto L_n^{- \frac{\ln \frac{b}{2}}{\ln 2} } \\
\sqrt{ \overline{ f_n^2} - (\overline{ f_n} )^2}
&& \propto L_n^{- \omega_W'(b) }
\ \ \ {\rm with } \ \ \ \omega_W'(b)= \frac{\ln \frac{b^2}{2}}{2 \ln 2}
\label{wettinghighexponents}
\end{eqnarray}

\subsection{ Low-temperature RG flow} 

In the low-temperature phase, the free-energies $f_n$ of Eq. \ref{defzwetting}
grow extensively with the length $L_n=2^n$,
and thus at large scale, the recursion is dominated by the first term in Eq.
\ref{zrecursionwetting} 
\begin{eqnarray}
f_{n+1} \opsimeq f_n^{(1)} +f_n^{(2)} +...
\label{recursionwettinglow}
\end{eqnarray}
The probability distribution of the free-energy thus converges to a
Gaussian, the average and the width grows as
\begin{eqnarray}
\overline{ f_n} \propto L_n \\
\sqrt{ \overline{ f_n^2} - (\overline{ f_n} )^2}
\propto L_n^{1/2 }
\label{wettinglowexponents}
\end{eqnarray}

\subsection{ Analysis of the critical invariant distribution} 

At criticality, to avoid the high-temperature and low-temperature described
above, the free-energy $f_n$ of Eq. \ref{defzwetting}
should remain a random variable
of order $O(1)$ with some scale-invariant probability distribution $P_c(f)$
defined on $]-\infty,0]$.
Equivalently, the variable $z_n=e^{- \beta_c f_n}$ should have a scale-invariant 
probability distribution $Q_c(z)$ defined on $[1,+\infty[$.
In the following, we derive some of their properties.

\subsubsection{ Left-tail behavior of the free-energy distribution}

Let us introduce the left-tail exponent $\eta_c$
\begin{eqnarray}
\ln P_c(f)  \opsimeq_{f \to -\infty} - \gamma (-f)^{\eta_c} +... 
\label{tailscritiwett}
\end{eqnarray}
where $(...)$ denote the subleading terms.

In the region where $f \to -\infty$, one has effectively
the low-temperature recursion 
\begin{eqnarray}
f \opsimeq f^{(1)} +f^{(2)} +...
\label{recursioncritileft}
\end{eqnarray}
A saddle-point analysis shows that if $f^{(1)}$ and $f^{(2)}$
have a probability distribution with the left tail given by
 Eq \ref{tailscritiwett},
their sum $f$ has for left tail 
$\ln P_c(f) \opsimeq - \gamma (-f)^{\eta_c} 2^{1-\eta_c} +...$.
The stability of the critical distribution thus fixes the value
of the left tail exponent to
\begin{eqnarray}
\eta_c=1
\label{etacwett}
\end{eqnarray}
So the distribution $P_c(f)$ decays exponentially
\begin{eqnarray}
P_c(f) \opsimeq_{f \to - \infty} e^{\gamma f} (...) 
\label{lefttailexpo}
\end{eqnarray}
This means that the corresponding distribution $Q_c(z)$
of $z=e^{-\beta_c f}$ presents a power-law tail 
\begin{eqnarray}
Q_c(z) \opsimeq_{z \to + \infty} \frac{ \Phi(\ln z) }{ z^{1+\mu} }
\label{powerlawqc}
\end{eqnarray}
with some exponent
\begin{eqnarray}
\mu= \frac{\gamma}{\beta_c}
\label{defimu}
\end{eqnarray}
and where $\Phi(\ln z)$ represents
the subleading terms.

\subsubsection{ Analysis in terms of multiplicative stochastic processes} 

\label{wettmsp}

The fact that a power-law appears in the stationary distribution of some random iteration
is reminiscent of multiplicative stochastic processes, whose main properties
are recalled in Appendix \ref{multiplicative}.
For a multiplicative stochastic process $X_n$ described by Eq. \ref{msp},
the stationary distribution presents 
a power-law tail of exponent $\mu$ that can be computed in terms
of the statistics of the random coefficient $a_n$ via Eq. \ref{mspmu}.
Here, for the quadratic renormalization of Eq. \ref{zrecursionwetting},
it is the process $z_n$ itself that also plays the role
of the random multiplicative coefficient.
As a consequence, it is instructive to analyse 
the recursion of Eq. \ref{zrecursionwetting}
along the same lines used to study multiplicative stochastic processes.

The necessary stability condition of Eq. \ref{stabcondition} translates here
into the following condition
\begin{eqnarray}
\overline{ \ln \frac{z}{b} } \equiv \int_1^{+\infty}  dz 
Q_c(z) \ln z - \ln b <0
\label{stabconditionbis}
\end{eqnarray}
The condition of Eq. \ref{mspmu} that ensures the stability of
the power-law tail via iteration
translates into the following self-consistent condition
 for the tail exponent $\mu$ introduced in Eq. \ref{powerlawqc}
\begin{eqnarray}
2 \overline{\left( \frac{z}{b} \right)^{\mu} }
\equiv 2 \int_1^{+\infty}  dz 
Q_c(z) \ \left( \frac{z}{b} \right)^{\mu} =1
\label{selfmu}
\end{eqnarray}
The argument is similar to the 
computation of Eqs \ref{eqmsp}-\ref{eqmspasymp}, 
the additional factor of $2$ coming from the fact that $z$ large
corresponds to either $z^{(1)}$ large or $z^{(2)}$ large.
The condition of Eq. \ref{selfmu} means
 in particular that the subleading term $\Phi(\ln z)$
in Eq. \ref{powerlawqc} should ensure the convergence 
at $(+ \infty)$ of the following integral
\begin{eqnarray}
 \overline{z^{\mu} } =
 \int^{+\infty}  dz 
Q_c(z) z^{\mu} \sim \int^{+\infty} \frac{dz}{z} \Phi(\ln z)
 \sim \int^{+\infty} dw \Phi(w) < +\infty
\label{subphi}
\end{eqnarray}
Since $\Phi$ is a subleading term in Eq  \ref{powerlawqc},
it should not contain an exponential,
so its decay should be a power-law
\begin{eqnarray}
 \Phi(w) \opsimeq_{w \to \infty} \frac{1}{ w^{1+\sigma} } 
\ \ \ \ {\rm with } \ \ \ \sigma>0
\label{subphipower}
\end{eqnarray}

Then moments of order $k \leq \mu$ are finite, whereas 
moments of order $k > \mu$ diverge
\begin{eqnarray}
 \int^{+\infty}  dz 
Q_c(z) z^{k} = + \infty \ \ {\rm for} \ \ \ \ \ \ \   k>\mu
\label{momentsdv}
\end{eqnarray}

In contrast with multiplicative stochastic processes where the condition of Eq. \ref{mspmu}
allows to compute the tail exponent $\mu$ in terms
of the known statistics of the random coefficient $a_n$,
we have obtained here only a self-consistent equation :
the selected tail exponent $\mu$ in the region $z \to \infty$
is the exponent that satisfies the condition of Eq. \ref{selfmu}
that involves the whole distribution for $z \in (1,+\infty[$. 
However, even if we cannot explicitly compute this exponent $\mu$,
we can try to locate it with respect to integer values
by considering the integer moments.

\subsubsection{  Reminder on transitions of integer moments } 

An important property of quadratic renormalizations
is that they lead to closed renormalizations for the integer moments.
We now briefly recall the behavior of the first moments discussed
in \cite{Der_wett}.
The closed recursion satisfied by the first moment \cite{Der_wett}
\begin{eqnarray}
\overline{z_{n+1}} = \frac{ (\overline{z_{n}})^2+b-1 }{b}
\label{recursionzav}
\end{eqnarray}
coincides with the pure case equation of Eq. \ref{mappingpurewetting}.
Using the initial condition of Eq. \ref{zrecursioninitial},
the unstable fixed point of Eq. \ref{rcpure}
allows the define the annealed temperature via
$\overline{ e^{ -\epsilon_i/T_{ann}}} = b-1$ :
for $T> T_{ann}$, the averaged value $ \overline{z_{n}}$ goes to $1$,
whereas for $T<T_{ann}$, the averaged value $ \overline{z_{n}}$ 
goes to $+\infty$. 
So $T_{ann}$ represents the transition of the first moment.
To locate $T_c$ with respect to $T_{ann}$,
we have to distinguish two possibilities

(i) if the tail exponent $\mu$ of Eq. \ref{powerlawqc} satisfies $0<\mu<1$,
then its first moment diverges at criticality $\overline{z_c}=+ \infty$
and we have the strict inequality $T_c<T_{ann}$.

(ii) if the tail exponent $\mu$ satisfies $\mu>1$,
then its first moment is finite at criticality. 
The only possible finite stable value is 
$\overline{z_c}=b-1$ and the critical temperature then
coincides with the annealed temperature  $T_c=T_{ann}$.
However, the analysis of the recursion for the variance
leads to the conclusion that the critical temperature 
is strictly lower than the annealed temperature $T_c<T_{ann}$ 
as soon as disorder is relevant 
$b \geq 2+\sqrt{2} \simeq 3.414$ \cite{Der_wett}.

In conclusion, whenever disorder is relevant at criticality,
one has the strict inequality $T_c<T_{ann}$, 
the first moment diverges $\overline z_c = +\infty$, 
and the tail exponent $\mu$
of Eq. \ref{powerlawqc} is smaller than $1$
\begin{eqnarray}
 0<\mu<1
\label{rangewettingmu}
\end{eqnarray}

\subsection{ Critical exponent $\nu$ }

\label{nuwetting}

\subsubsection{  Equivalence with a directed polymer on a Cayley tree }

In the pure case, the critical exponents
are obtained from the linearization around the fixed point (see
section \ref{purewetting}).
To follow the same strategy in the disordered case, we set 
$z_n=z_c+\delta_n$. At linear order, we obtain the recursion
\begin{eqnarray}
\delta_{n+1} \simeq \frac{z_c^{(1)}}{b } \delta_{n}^{(2)}
+   \frac{z_c^{(2)}}{b } \delta_{n}^{(1)}
\label{epswetting}
\end{eqnarray}
where $z_c^{(1,2)}$ are distributed with the critical distribution $Q_c(z)$.
As in the pure case, it is also interesting to write the recursion 
for the energy $E_n$ 
\begin{eqnarray}
E_{n+1} \simeq \frac{ z_c^{(1)} z_c^{(2)}
 (E_n^{(1)}+E_n^{(2)}) }{z_c^{(1)}z_c^{(2)} + b-1 }
\label{enerwetting}
\end{eqnarray}
so that the combination $U_n \equiv z_n E_n$  satisfies 
at criticality the same recursion as in Eq. \ref{epswetting}
\begin{eqnarray}
U_{n+1} = \frac{z_c^{(1)}}{b } U_{n}^{(2)}
+   \frac{z_c^{(2)}}{b } U_{n}^{(1)}
\label{unwetting}
\end{eqnarray}

The recurrence of Eq \ref{epswetting}
 coincides with the transfer matrix 
\begin{eqnarray}
Z_{L+1}= \sum_{i=1}^K e^{- \epsilon_i} Z_L^{(i)} 
\label{defcayley}
\end{eqnarray}
for the partition function $Z_L$
of a directed polymer on a Cayley tree 
of branching ratio $K=2$ with random bond energies $\epsilon_i$.
 \cite{Der_Spo,Der_review}.
The differences with Eq. \ref{epswetting}
we are interested in are the following :

(i) the partition function $Z_L$ in Eq \ref{defcayley}
is positive by definition,
whereas here the random perturbation $\delta_n$ in Eq.
\ref{epswetting} are a priori of arbitrary sign.
Eq. \ref{epswetting} is thus more related to the case
of a  directed polymer model with complex weights studied in
\cite{cayleycomplex}.

(ii) the weights $ e^{- \beta \epsilon_i}$
associated to the bond energies $\epsilon_i$ in Eq. \ref{defcayley}
are now random weights $\frac{z_c}{b}$
distributed with the fixed point distribution $Q_c(z)$
\begin{eqnarray}
e^{- \beta \epsilon_i} \to  \frac{z_c}{b}
\label{equiwetting}
\end{eqnarray}
In particular, these weights present a broad power-law tail
in $1/z_c^{1+\mu}$ in contrast with the usual case
where the energies $\epsilon_i$ are Gaussian.

The difference (ii) turns out to be very important
as we now explain.

\subsubsection{  Tails analysis }

Let us consider the first iteration 
\begin{eqnarray}
\delta_1 = \frac{z_c^{(1)}}{b } \delta_{0}^{(2)}
+   \frac{z_c^{(2)}}{b } \delta_{0}^{(1)}
\end{eqnarray}
Suppose we start with a narrow distribution ${\cal P}_0(\delta_0)$
for the random initial perturbation $\delta_0$.
The distribution ${\cal P}_1(\delta_1) $
after the first iteration will nevertheless present 
power-law tails inherited from the fixed point distribution
 $Q_c(z_c) \sim \Phi(\ln z_c) /z_c^{1+\mu}$
with $0<\mu<1$ (Eqs \ref{powerlawqc} and \ref{rangewettingmu}).
More precisely, the tail in the region $\delta_1 \to +\infty$ 
is dominated by the events where $z_c^{(1)}$ is large with 
$\delta_{0}^{(2)} >0$ or where $z_c^{(2)}$ is large with 
$\delta_{0}^{(1)} >0$, and one obtains
\begin{eqnarray}
{\cal P}_1 (\delta_1) &&  \opsimeq_{\delta_1 \to +\infty}
2 \int dz_c Q_c(z_c) \int_0^{+\infty} d \delta_0 {\cal P}_0(\delta_0)
\delta \left[ \delta_1- \frac{z_c}{b } \delta_{0} \right] \\
&& \opsimeq_{\delta_1 \to +\infty} 
\frac{\Phi(\ln \delta_1)}{\delta_1^{1+\mu}} \left[ \frac{2}{b^{\mu}}
\int_0^{+\infty} d \delta_0 {\cal P}_0(\delta_0)\ \delta_0^{\mu} \right]
\end{eqnarray}
Similarly, the left tail reads
\begin{eqnarray}
{\cal P}_1 (\delta_1) 
&& \opsimeq_{\delta_1 \to -\infty} 
\frac{\Phi(\ln \vert \delta_1 \vert)}{\vert \delta_1 \vert^{1+\mu}}
 \left[ \frac{2}{b^{\mu}}
\int_{-\infty}^{0} d \delta_0 {\cal P}_0(\delta_0)\ 
\vert \delta_0\vert^{\mu} \right]
\end{eqnarray}

It is then clear that by iteration all distributions ${\cal P}_n(\delta_n) $
will present these power-law tails
\begin{eqnarray}
{\cal P}_n (\delta_n) 
&& \oppropto_{\delta_n \to \pm \infty} 
\frac{\Phi(\ln \vert \delta_n \vert)}{\vert \delta_n \vert^{1+\mu}}
\label{defamplitude}
\end{eqnarray}
Since we are looking for the Lyapunov exponent $v$
governing the typical growth of the perturbation
\begin{eqnarray}
\frac{\delta_{n+1}}{\delta_n} \sim e^{v }
\label{defvwett}
\end{eqnarray}
it is convenient to rescale the iteration of Eq. 
\ref{epswetting} by the factor $e^{-v}$
\begin{eqnarray}
y_{n+1} = e^{-v} \left[ \frac{z_c^{(1)}}{b } y_{n}^{(2)}
+   \frac{z_c^{(2)}}{b } y_{n}^{(1)} \right]
\label{epswettingrescal}
\end{eqnarray}
and to ask that the probability distribution $P_n(y)$
converges as $n \to \infty$
towards a stable distribution $P_{\infty}(y)$
presenting the tails (Eq \ref{defamplitude})
\begin{eqnarray}
P_{\infty} (y) 
&& \opsimeq_{y \to \pm \infty} B^{\pm}
\frac{\Phi(\ln \vert y \vert)}{\vert y \vert^{1+\mu}}
\label{defamplitudey}
\end{eqnarray}
Reasoning as before, a large value of $y_{n+1}$ corresponds to a large value
of one of the four variables $(y_{n}^{(1)},y_{n}^{(2)},z_c^{(1)},z_c^{(2)}) $,
and one obtains the following equations
\begin{eqnarray}
B^+  && = \left[  B^+  \frac{2 e^{-\mu v} \overline{ z_c^{\mu} } }{b^{\mu}}
+ \frac{2 e^{-\mu v}}{b^{\mu}}
\int_0^{+\infty} dy P_{\infty}(y) \ y^{\mu}  \right] \\
B^-  && = \left[  B^-  \frac{2 e^{-\mu v} \overline{ z_c^{\mu} } }{b^{\mu}}
+ \frac{2 e^{-\mu v}}{b^{\mu}}
\int_{-\infty}^0 dy P_{\infty}(y) \vert y \vert^{\mu}  \right]
\end{eqnarray}
This shows that positive perturbations (initial distribution
${\cal P}_0(\delta_0<0)=0$) or symmetric perturbations 
(symmetric initial distribution
${\cal P}_0(\delta_0)={\cal P}_0(-\delta_0)$) actually lead to
the same Lyapunov exponent $v$
\begin{eqnarray}
e^{\mu v}  =  \frac{2  \overline{ z_c^{\mu} } }{b^{\mu}}
+ \frac{2 }{b^{\mu}}  \int_0^{+\infty} dy \frac{P_{\infty}(y)}{B_+} \ y^{\mu}
= 1+ 
\frac{2 }{b^{\mu}}  \int_0^{+\infty} dy \frac{P_{\infty}(y)}{B_+} \ y^{\mu} 
\label{resvelocitywett}
\end{eqnarray}
where we have used Eq. \ref{selfmu}.
The first term corresponds to the usual term 
for the velocity of the travelling wave approach \cite{Der_Spo,Der_review},
whereas the second term originates from the broad distribution
of weights $(z_c/b)$.
Its physical meaning is the following :
in the usual case of a narrow distribution of the weights,
the travelling wave approach allows to compute the velocity
in terms of the weight statistics alone, because one can write
a closed equation for the tail of the process \cite{Der_Spo,Der_review};
in the present case, the tail of the process does not satisfy a closed
equation, because the broadness of the weight distribution
induces some interaction between the tail and the bulk
of the process : the second term in Eq. \ref{resvelocitywett}
represents the influence of the bulk of the distribution $P_{\infty}(y)$
onto the tail of exponent $\mu$.

The exponent $\nu$ describing the power-law growth $\delta_n \sim L_n^{1/\nu}
\sim e^{vn}$ reads in terms of the Lyapunov exponent
\begin{eqnarray}
\nu= \frac{\ln 2}{v} 
\label{nuwett}
\end{eqnarray}
Note that the presence of the second term in Eq. \ref{resvelocitywett}
is crucial to obtain a finite exponent $\nu$ :
without this second term, the Lyapunov exponent $v$
would vanish ($v=0$) and the correlation length
exponent would diverge ($\nu=\infty$).

\section{ Numerical study of the wetting transition }

\label{secwettingnume}

\subsection{  Numerical method }

\label{wettnume}

We have performed numerical simulations with the so-called 'pool-method'
which is very much used for disordered systems on hierarchical lattices
\cite{hierarchicalspinglass,Coo_Der} : 
the idea is to represent the probability distribution
$P_n(F_n)$ of the free-energy $F_n=-T  \ln Z_n$ at generation
$n$, by a pool of $N$ values $\{F_n^{(1)},..,F_n^{(N)} \}$.
The pool at generation $(n+1)$ is then obtained as follows :
each value $F_{n+1}^{(i)}$ is obtained by choosing 
two values at random from the pool of generation $n$ and by applying
the renormalization Eq. \ref{zrecursionwetting}.

The results presented in this Section have been obtained
for the branching ratio $b=5$, with a pool number $N=4.10^7$,
with initial Gaussian energies (Eq. \ref{gaussian}).
The corresponding annealed temperature is 
\begin{eqnarray}
T_{ann} = \frac{1}{\sqrt{2 \ln(b-1) } } \simeq 0.60055
\label{tannealed}
\end{eqnarray}

Finally, the relation between the true free-energy $F_n=- T \ln Z_n$
and the reduced free-energy $f_n=-T \ln z_n$ used in the previous section
is simply (Eq \ref{defzwetting})
\begin{eqnarray}
F_n= f_n-T \ln Y_n
\label{grandfpetitf}
\end{eqnarray}
where $Y_n=b^{L_n-1}$  does not contain any disorder.
As a consequence, the two free-energy distributions
have the same width $\Delta F_n = \Delta f_n$, and the same tail properties.

\subsection{ Flow of the free-energy width $\Delta F_L$ } 

\begin{figure}[htbp]
%\begin{figure}
\includegraphics[height=6cm]{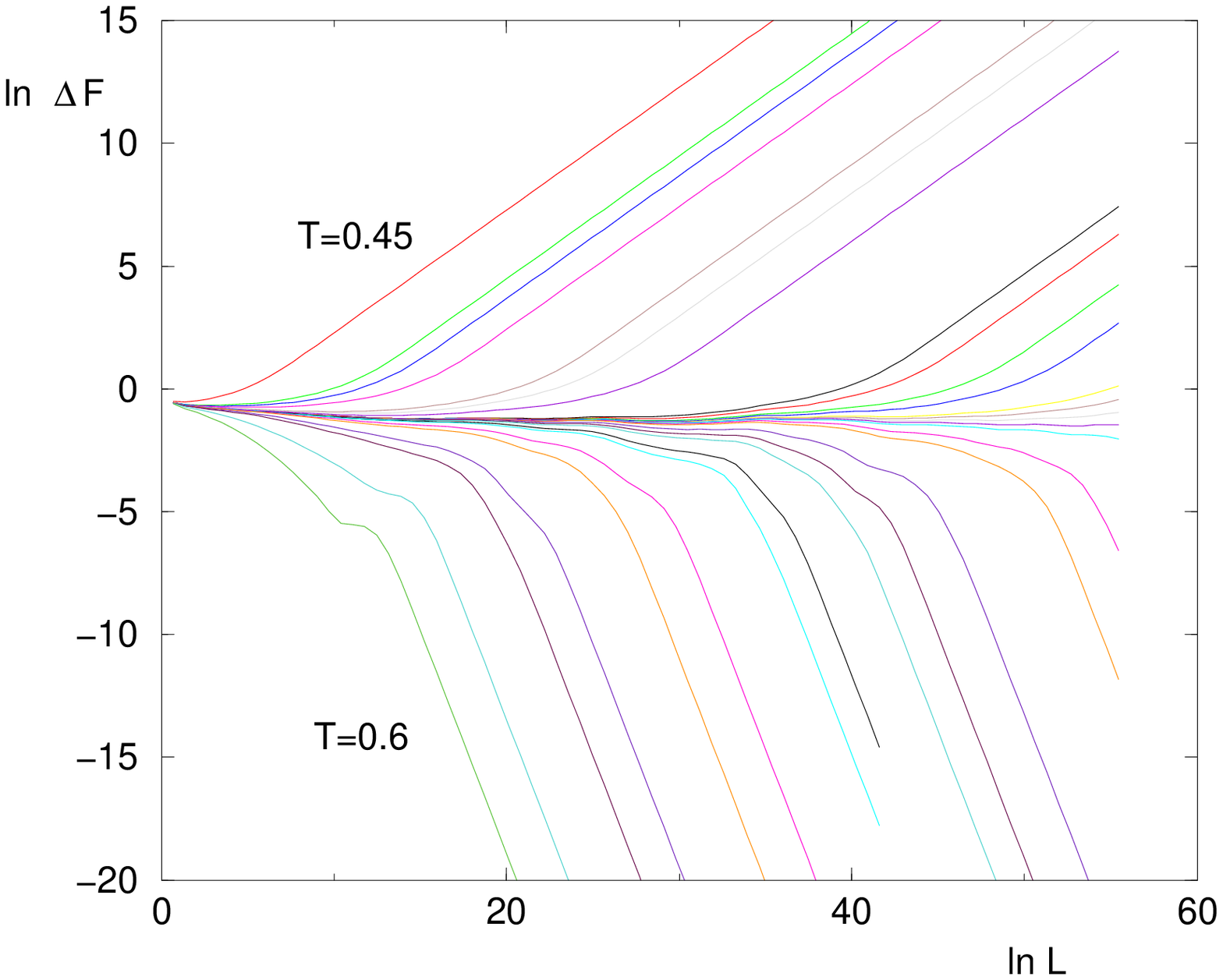}
\hspace{1cm}
\caption{(Color online) Wetting transition :
log-log plot of the width $\Delta F(L)$
of the free-energy distribution as a function of $L$, for many temperatures.   }
\label{figwettb5freewidth}
\end{figure}

The flow of the free-energy width $\Delta F_L$ as $L$ grows
is shown on Fig. \ref{figwettb5freewidth} for many temperatures.
One clearly sees the two attractive fixed points on this log-log plot.

For $T>T_c$, the free-energy width decays asymptotically 
with the exponent $\omega_W'(b)$ introduced in Eq. \ref{wettinghighexponents}
\begin{eqnarray}
\Delta F(L) \simeq \left( \frac{L}{\xi_F^+(T)} \right)^{- \omega_W'(b=5)}
 \ \ { \rm with } \ \ \omega_W'(b=5)=
 \frac{\ln (b^2/2) }{ 2 \ln 2} = 1.8219..
\label{fwidthabovewett}
\end{eqnarray}
where $\xi_F^+(T)$ is the corresponding correlation length
that diverges as $T \to T_c^+$.

For $T<T_c$, the free-energy width grows asymptotically 
with the exponent $1/2$ (see Eq. \ref{wettinglowexponents} )
\begin{eqnarray}
\Delta F(L) \simeq \left( \frac{L}{\xi_F^-(T)} \right)^{ 1/2}
\label{fwidthbelowwett}
\end{eqnarray}
where $\xi_F^-(T)$ is the corresponding correlation length
that diverges as $T \to T_c^-$.

The critical temperature obtained by this pool method depends
of the pool, i.e. of the discrete sampling with $N$ values
of the continuous probability distribution. 
It is expected to converge towards the thermodynamic
critical temperature $T_c$ only in the limit $N \to \infty$.
Nevertheless, for each given pool,
the flow of free-energy width allows a very precise
determination of this pool-dependent critical temperature,
for instance in the case considered 
$0.52415 < T_c^{pool} < 0.52416$,
which is significantly below the annealed temperature of Eq \ref{tannealed}.

\subsection{ Divergence of the correlation lengths $\xi_F^{\pm}(T)$ } 

\begin{figure}[htbp]
%\begin{figure}
\includegraphics[height=6cm]{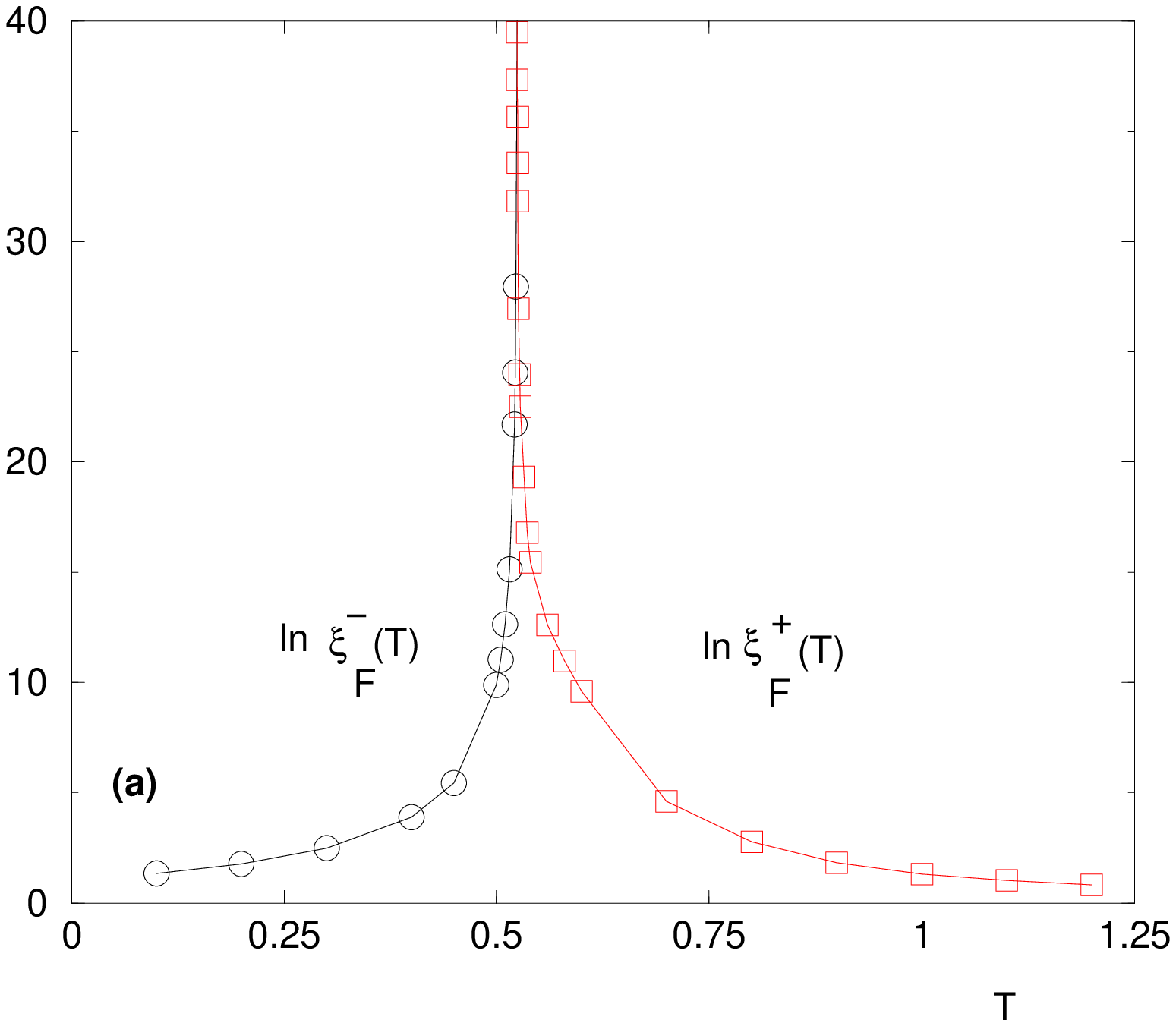}
\hspace{1cm}
\includegraphics[height=6cm]{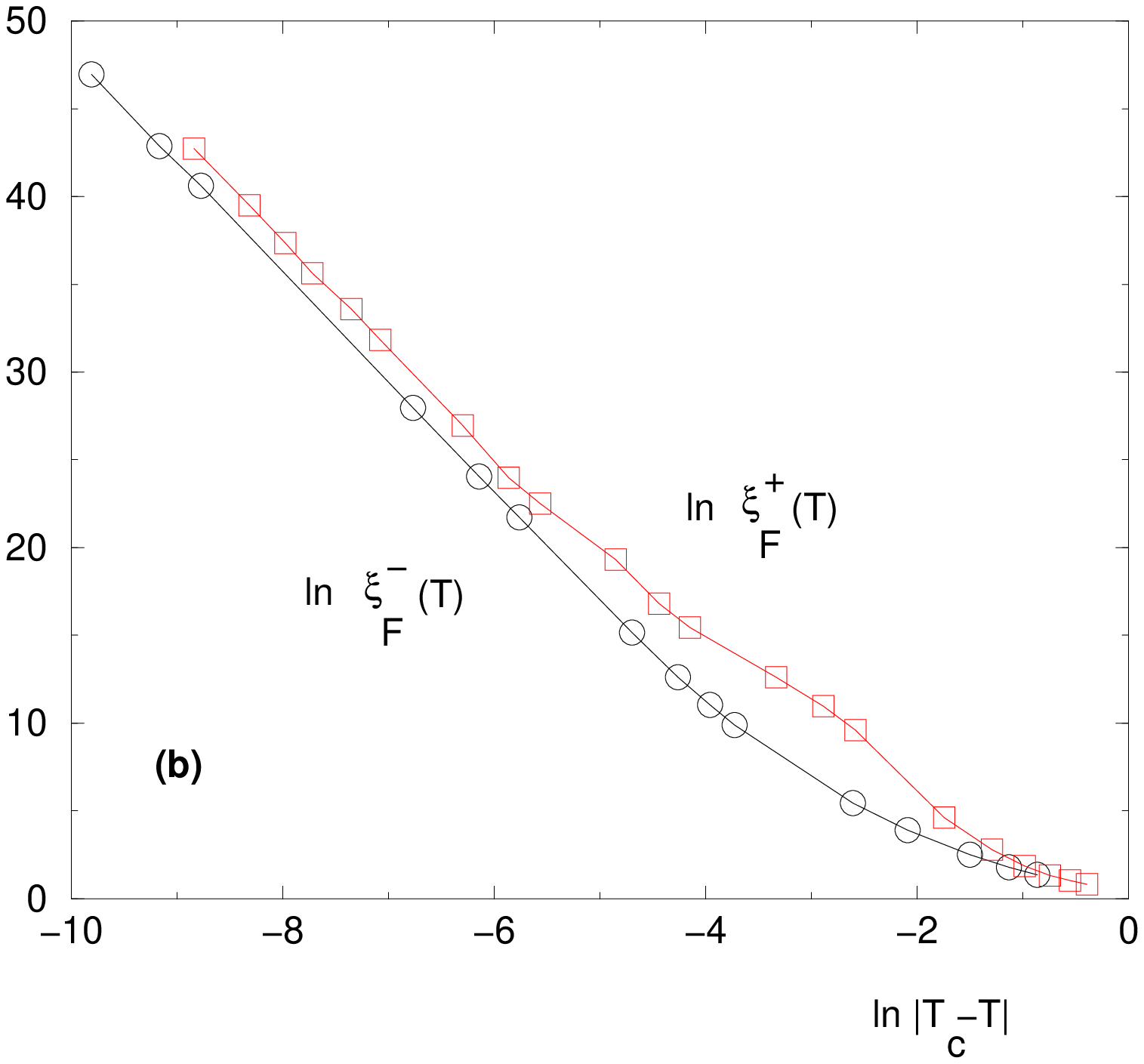}
\caption{(Color online) Wetting transition :
Correlation length $\xi_F^{\pm}(T)$ as measured from the 
behavior of the free-energy width (Eqs \ref{fwidthabovewett} and
\ref{fwidthbelowwett})
(a) $\ln \xi_F^{\pm}(T)$ as a function of $T$
(b) $\ln \xi_F^{\pm}(T)$ as a function of 
$\ln \vert T_c-T \vert$  :
the asymptotic slopes are of order $\nu \sim 6.2 $
}
\label{figb5logxifreewett}
\end{figure}

The correlation lengths $\xi_F^{\pm}(T)$ as measured from the free-energy
width asymptotic behaviors above and below $T_c$ (Eqs \ref{fwidthabovewett} 
and \ref{fwidthbelowwett} ) are shown on Fig. \ref{figb5logxifreewett} (a).
The plot in terms of the variable $\ln \vert T_c-T \vert$
shown on Fig. \ref{figb5logxifreewett} (b) indicate a power-law divergence
with the same exponent 
\begin{eqnarray}
\xi_F^{\pm}(T) \oppropto_{T \to T_c} \vert T-T_c \vert^{-\nu} 
\ \ { \rm with } \ \ \nu \simeq 6.2
\label{xifreenuwett}
\end{eqnarray}

\subsection{ Histogram of the free-energy }

\begin{figure}[htbp]
%\begin{figure}
%\includegraphics[height=6cm]{xm51wett.mchistob5.tall.eps}
%\hspace{1cm}
\includegraphics[height=6cm]{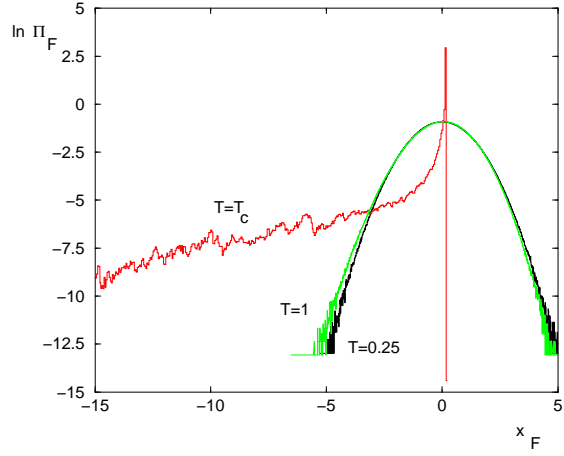}
\caption{(Color online) Wetting transition: Log-plot of the asymptotic
distribution $\Pi_F$ of the rescaled free-energy  
$x_F= \frac{F-F_{av(L)}}{\Delta F(L)} $  
in the low-temperature phase (here $T=0.25$), in the high-temperature phase
( here $T=1$) and at criticality (here $T_c^{pool}=0.524155$) }
\label{figwettb5freehisto}
\end{figure}

The asymptotic probability distribution $\Pi_F$ of
the rescaled free-energy 
\begin{eqnarray}
x_F \equiv \frac{F-F_{av(L)}}{\Delta F(L)}
\end{eqnarray}
is shown on Fig \ref{figwettb5freehisto} for three temperatures :

(i) the distribution is Gaussian both for $T>T_c$
and $T<T_c$ as expected from Eqs \ref{recursionwettinghigh} 
and \ref{recursionwettinglow}.

(ii) at criticality, one clearly see that a left-tail 
develops in the region $f \to -\infty$
with tail exponent $\eta_c =1$ in agreement with Eq \ref{etacwett}.
The corresponding power-law exponent of Eq \ref{defimu}
of the fixed-point distribution of Eq. \ref{powerlawqc}
is of order 
\begin{eqnarray}
\mu \sim 0.45
\label{muvaluewett}
\end{eqnarray}
The measure is not very precise because
one clearly sees on Fig. \ref{figwettb5freehisto}
that on top of this power-law, there exists oscillations
reflecting the discrete nature of the renormalization.
But this value is anyway in the expected interval 
of Eq \ref{rangewettingmu}.

\subsection{ Flow of the energy width  } 

\begin{figure}[htbp]
%\begin{figure}
\includegraphics[height=6cm]{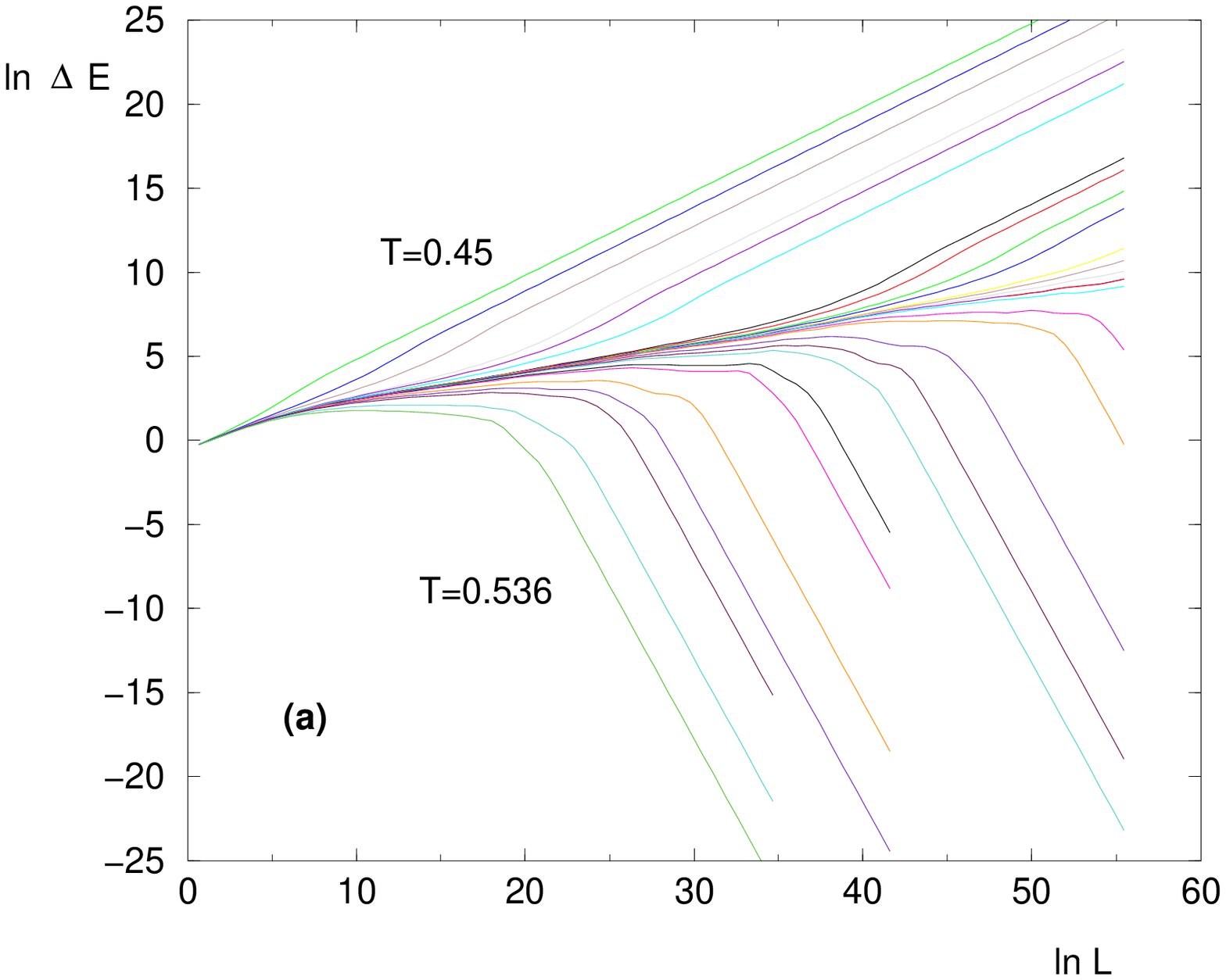}
\hspace{1cm}
 \includegraphics[height=6cm]{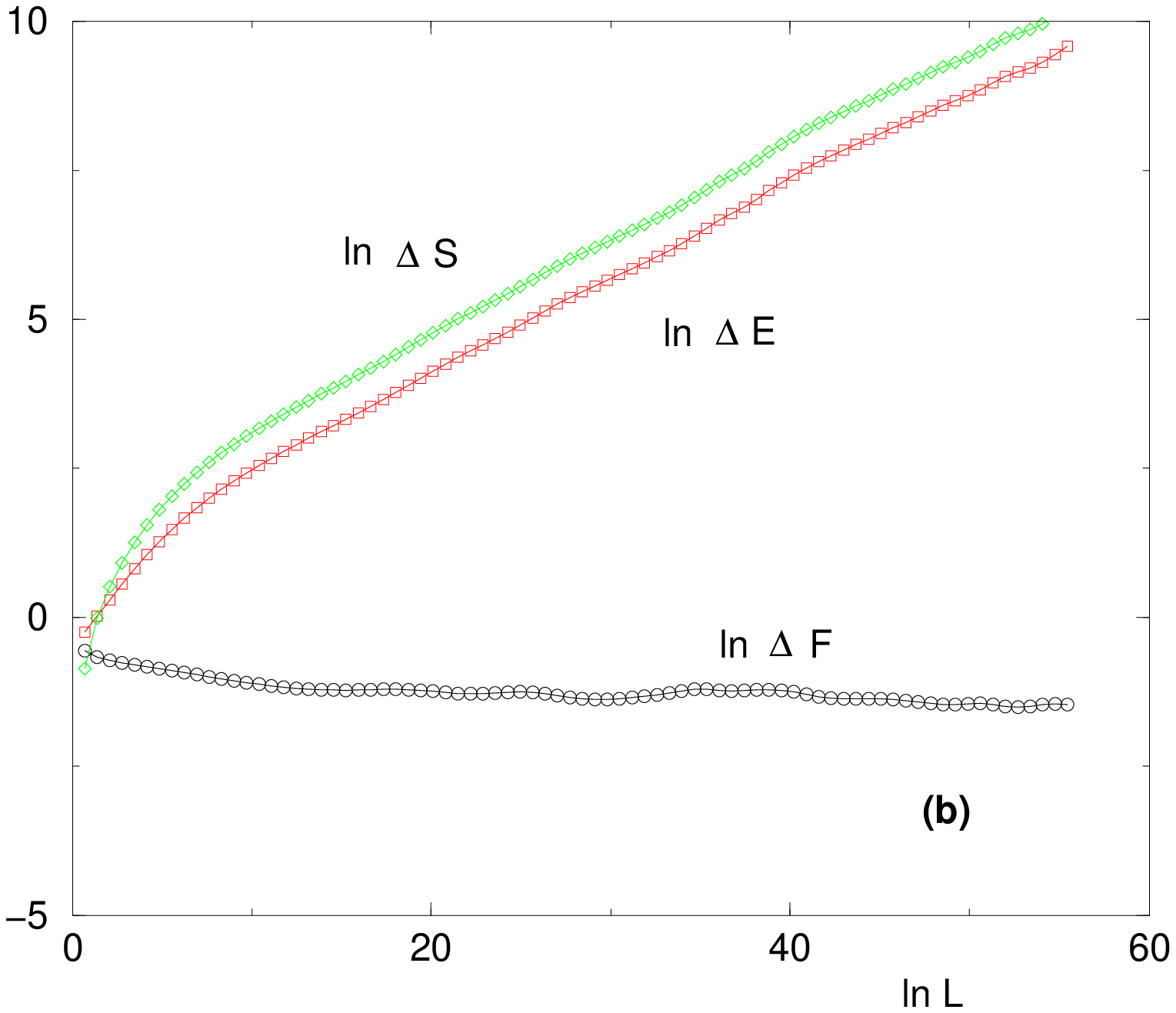}
\caption{(Color online) Wetting transition :
Flow of the widths $\Delta E(L)$ 
of the energy distribution as $L$ grows
(a) $\ln \Delta E(L)$ as a function of $\ln L$ for many temperatures 
(b) Comparison
of $\ln \Delta E(L)$ , $ \ln \Delta S(L)$ and $ \ln \Delta F(L)$
as a function of $\ln L$ at criticality ($T_c^{pool}=0.524155$).}
\label{figwettb5enerwidth}
\end{figure}

The flow of the energy width $\Delta E(L)$ 
as $L$ grows
are shown on Fig. \ref{figwettb5enerwidth} for many temperatures.
For $T>T_c$, we find that the width decays asymptotically 
with the same exponent $\omega_{\infty}^{W}(b)$ as the free-energy
 (Eq \ref{fwidthabovewett})
\begin{eqnarray}
\Delta E(L) \simeq L^{- \omega_{\infty}^{W}(b)} 
\ \ { \rm with } \ \ \omega_{\infty}^{W}(b=5)=
 \frac{\ln (b^2/2) }{ 2 \ln 2} = 1.8219..
\label{wettewidthabove}
\end{eqnarray}
For $T<T_c$, 
this width  grows asymptotically 
with the exponent $1/2$ as the free-energy (Eq \ref{fwidthbelowwett})
\begin{eqnarray}
\Delta E(L) \simeq L^{\frac{1}{2}} \\
\label{wettewidthbelow}
\end{eqnarray}

Exactly at criticality,
the free-energy $\Delta F(L)$ width converges towards a constant,
whereas the energy width grows as a power-law 
(see Fig \ref{figwettb5enerwidth} b)
\begin{eqnarray}
\Delta E(L) \simeq L^{y_c} \ \ \ {\rm with } \ \ y_c \simeq 0.16
\label{wettewidthcriti}
\end{eqnarray}
This exponent is in agreement with the finite-size scaling relation $y_c=1/\nu$
with $\nu \simeq 6.2$ (see Eqs \ref{xifreenuwett})

\section{ Directed polymer model on diamond hierarchical lattices}

\label{secdp}

In this Section we study the directed polymer model, whose partition function
satisfies the quadratic renormalization of Eq \ref{zrecursion}.
In contrast with the wetting case described above, the transition only
exists in the presence of disorder.
Since we are interested into the asymptotic distribution of the free-energy,
it is convenient to rewrite the free-energy $F_n^{(a)}$ of a sample $(a)$
of generation $n$ as
\begin{eqnarray}
 F_n^{(a)} \equiv \ln Z_n^{(a)} =  \overline{ F_n}+ \Delta_n u_a
\label{scalingfree}
\end{eqnarray}
where $u_a$ is a random variable of zero mean and width unity 
\begin{eqnarray}
\overline{u_a^2}=1
\label{unorma}
\end{eqnarray}
So $\Delta_n$ represents the width 
\begin{eqnarray}
\Delta_n= \left( \overline{ F_n^2} - (\overline{ F_n})^2 \right)^{1/2}
\label{defdeltan}
\end{eqnarray}

\subsection{ Low-temperature RG flow} 

In the low-temperature phase, the width $\Delta_n$ 
of the free-energy distribution grows with $n$.
So at large scale, the recursion is dominated by the maximal term in Eq.
\ref{zrecursion} 
\begin{eqnarray}
F_{n+1} \opsimeq min_{1 \leq a \leq b}
 \left( F_n^{(2a-1)} + F_n^{(2a)} \right)
\label{recursionDPlow}
\end{eqnarray}
This effective low-temperature recursion coincides with the recursion of the energy $E_0$
of the ground state studied in \cite{Der_Gri,Coo_Der}.
The whole low-temperature phase is thus described by
the zero-temperature fixed-point. In particular, the width
of the free-energy distribution grows as
\begin{eqnarray}
\Delta_n \opsimeq L_n^{\omega_0(b)}
\label{widthDPlow}
\end{eqnarray}
where $\omega_0(b)$ is the exponent governing the width 
of the ground-state energy $\Delta E_0 \sim L_n^{\omega_0(b)} $ 
studied in \cite{Der_Gri}.

\subsection{ High-temperature RG flow} 

In the high-temperature phase, the width $\Delta_n$ of the free-energy
distribution is expected to decay to zero.
The linearization in $\Delta_n$ of the recursion of Eq. \ref{zrecursion}
yields
\begin{eqnarray}
 \beta F_{n+1} && = - \ln \left[  \sum_{a=1}^b
 e^{- \beta \left( F_n^{(2a-1)}+  F_n^{(2a)}\right) } \right]
 = 2 \beta  \overline{ F_n } 
 - \ln \left[ b - \beta \Delta_n \sum_{a=1}^b (u_{2a-1}+u_{2a})
 + O(\beta^2 \Delta_n^2)\right] \\
&& =  2 \beta  \gamma_n - \ln (b) 
 + \frac{ \beta \Delta_n }{b}  \sum_{a=1}^b (u_{2a-1}+u_{2a})
 + O(\beta^2 \Delta_n^2)
\label{frecursionhigh}
\end{eqnarray}
The consistence with the scaling form of Eq. \ref{scalingfree}
at generation $(n+1)$
\begin{eqnarray}
 F_{n+1} = \overline{F_{n+1}} + \Delta_{n+1} u
\label{scalinghighbis}
\end{eqnarray}
yields 
\begin{eqnarray}
\overline{F_{n+1}} && =   2 \overline{F_{n}} - T \ln (b) \nonumber \\
\Delta_{n+1} u && = \frac{ \Delta_n }{b}  \sum_{a=1}^b (u_{2a-1}+u_{2a})
\label{recurrencehigh}
\end{eqnarray}
The normalization condition of Eq. \ref{unorma} yields
\begin{eqnarray}
\Delta_{n+1}  = \sqrt{ \frac{2}{b} } \Delta_n 
\label{widthhigh}
\end{eqnarray}
and 
\begin{eqnarray}
u = \frac{1}{\sqrt{2b}}  \sum_{a=1}^b (u_{2a-1}+u_{2a})
\label{uhigh}
\end{eqnarray}

For $b>2$, this high-temperature phase exists,
the probability distribution of the free-energy converges to a
Gaussian. The width decays as
the following power-law of the length $L_n=2^n$
\begin{eqnarray}
\Delta_n
\propto L_n^{-  \omega_{\infty}(b)} \ \ \ {\rm with} \ \ \ \ \ 
\omega_{\infty}(b)=\frac{\ln \frac{b}{2}}{2 \ln 2} 
\label{fwidthhighdp}
\end{eqnarray}
In this regime, the rescaled variable $u$ evolves according to Eq. \ref{uhigh}
and thus becomes Gaussian upon iteration.

\subsection{ Analysis of the critical point} 

At criticality, to avoid the high-temperature and low-temperature described
above, the width $\Delta_n$ should converge as $n \to \infty$
towards a finite value $\Delta_c$.
In particular, the fluctuating part of the free-energy
\begin{eqnarray}
f_n^{(a)} \equiv F_n^{(a)} - \overline{ F_n}  
= \Delta_c u^{(a)}
\label{fluctuatingfree}
\end{eqnarray}
should remain a random variable of order $O(1)$, of zero mean,  distributed
 with some scale-invariant probability distribution $P_c(f)$
defined on $]-\infty,+\infty[$.
Equivalently, the variable
\begin{eqnarray}
z_n^{(a)} \equiv e^{- \beta_c f_n^{(a)} } =  e^{- \beta_c \Delta_c u^{(a)}}
\label{fluctuatingz}
\end{eqnarray}
should have a scale-invariant 
probability distribution $Q_c(z)$ defined on $[0,+\infty[$,
with
\begin{eqnarray}
\overline{ \ln z} = \int_0^{+\infty} dz \ln z Q_c(z)=0
\label{conditionlnz}
\end{eqnarray}
The recursion for these variables $z_n$ reads
\begin{eqnarray}
z_{n+1} = \frac{1}{\cal B} 
\sum_{a=1}^b z_n^{(2a-1)} z_n^{(2a)}
\label{zreducedrecursion}
\end{eqnarray}
where
\begin{eqnarray}
{\cal B} \equiv \lim_{n \to \infty} \left( \overline{ F_{n+1} }- 2 \overline{ F_n} \right)
\label{defcalB}
\end{eqnarray}
should be finite ( otherwise the recursion
of Eq. \ref{zreducedrecursion} would not lead to a
 non-trivial stationary distribution $Q_c(z)$).

\subsubsection{ Left-tail behavior of the free-energy distribution}

Let us introduce the left-tail exponent $\eta_c$
\begin{eqnarray}
\ln P_c(f)  \opsimeq_{f \to -\infty} - \gamma (-f)^{\eta_c} +... 
\label{tailscritidp}
\end{eqnarray}
where $(...)$ denote the subleading terms.

In the region where $f \to -\infty$, one has effectively
the low-temperature recursion of Eq. \ref{recursionDPlow}
\begin{eqnarray}
f \opsimeq min_{1 \leq a \leq b} \left( f^{(2a-1)} + f^{(2a)} \right)
\label{recursioncritileftdp}
\end{eqnarray}
As in the wetting case, 
a saddle-point analysis shows that 
the only stable left tail exponent is
\begin{eqnarray}
\eta_c=1
\label{etacdp}
\end{eqnarray}
So the distribution $P_c(f)$ decays exponentially
\begin{eqnarray}
P_c(f) \opsimeq_{f \to - \infty} e^{\gamma f} (...) 
\label{lefttailexpodp}
\end{eqnarray}
This means that the corresponding distribution $Q_c(z)$
of $z=e^{-\beta_c f}$ presents a power-law tail 
\begin{eqnarray}
Q_c(z) \opsimeq_{z \to + \infty} \frac{ \Phi(\ln z) }{ z^{1+\mu} }
\label{powerlawqcdp}
\end{eqnarray}
with some exponent
\begin{eqnarray}
\mu= \frac{\gamma}{\beta_c}
\label{defimudp}
\end{eqnarray}
and where $\Phi(\ln z)$ represents
the subleading terms.

The first moment is fixed by the recursion of Eq \ref{zreducedrecursion}
\begin{eqnarray}
\overline{ z} = \frac{ \cal B}{b}
\label{averagez}
\end{eqnarray}
As a consequence, the exponent $\mu$ of the power-law of Eq \ref{powerlawqcdp}
should satisfy
\begin{eqnarray}
\mu >1
\label{mubigger1}
\end{eqnarray}
and the parameter ${\cal B}$ representing the correction to extensivity (Eq \ref{defcalB})
is determined by
\begin{eqnarray}
{ \cal B} = b \overline{ z} =  b \int dz z Q_c(z)
\label{calB}
\end{eqnarray}
in terms of the fixed point distribution $Q_c(z)$.

\subsubsection{ Analysis in terms of multiplicative stochastic processes}

Again, as explained in section \ref{wettmsp}, it is instructive to analyse
 the recursion of Eq. \ref{zreducedrecursion}
from the point of view of multiplicative stochastic processes
(see Appendix \ref{multiplicative}).
The condition of Eq. \ref{stabcondition} translates here
into the following condition using Eq. \ref{conditionlnz}
\begin{eqnarray}
\overline{ \ln \frac{z}{{\cal B}} } = - \ln {\cal B } 
= - \ln ( b \overline{ z} ) <0
\label{stabconditiondp}
\end{eqnarray}
The condition of Eq. \ref{mspmu} translates into the following 
condition for the exponent $\mu$ introduced in Eq. \ref{powerlawqcdp}
\begin{eqnarray}
2 b \overline{\left( \frac{z}{\cal B} \right)^{\mu} }
= \frac{2}{b^{\mu-1}} 
\ \frac{ \overline{z^{\mu} }}{ (\overline{z})^{\mu}}
\equiv  \frac{2}{b^{\mu-1}}
 \frac{ \int_1^{+\infty}  dz Q_c(z) z^{\mu} }{  \left[ \int_1^{+\infty}  dz 
Q_c(z) z \right]^{\mu} }   =1
\label{selfmudp}
\end{eqnarray}
The argument is similar to Eqs \ref{eqmsp}-\ref{eqmspasymp}, 
the additional factor of $2b$ coming from the fact that $z$ large
corresponds to one of the $z^{(i)}$ being large.
This condition means in particular that the subleading term $\Phi(z)$
in Eq. \ref{powerlawqc} should ensure the convergence 
at $(+ \infty)$ as in the wetting case 
(see Eqs \ref{subphi} and \ref{momentsdv}).

\subsubsection{  Reminder on transitions of integer moments }

Let us now briefly recall the behaviors of the first moments discussed
in \cite{Coo_Der}. From Eq \ref{zrecursion}, one obtains \cite{Coo_Der}
\begin{eqnarray}
\overline{ Z_{n+1} } && = b \left( \overline{Z_n} \right)^2 \\
\overline{ Z_{n+1}^2 } && = b \left( \overline{Z_n^2 } \right)^2
+ b (b-1) \left( \overline{Z_n} \right)^4
\label{momentsdp}
\end{eqnarray}
so that the ratio of the moments of the rescaled variable $z$ defined in Eq \ref{fluctuatingz}
\begin{eqnarray}
r_2(n) \equiv \frac{\overline{ Z_{n}^2 }}{\left( \overline{Z_n} \right)^2} =
 \frac{\overline{ z_{n}^2 }}{\left( \overline{z_n} \right)^2} 
\label{defr2dp}
\end{eqnarray}
follows the closed recursion \cite{Coo_Der}
\begin{eqnarray}
r_2(n+1) = \frac{ r_2^2(n) +b-1 }{b}
\label{recr2dp}
\end{eqnarray}
that actually coincides with Eq. \ref{mappingpurewetting}
 for the pure wetting model.
The repulsive fixed point $r^*=b-1$ allows to define the temperature $T_2$
via $r_2(n=0)=b-1$ \cite{Coo_Der}:
   for $T>T_2$, the ratio $r_2(n)$ flows to $1$,
whereas for 
$T>T_2$, the ratio $r_2(n)$ flows to $(+\infty)$.
Similarly, the RG flow of ratios corresponding to higher moments
have been studied in \cite{Coo_Der}, with the conclusion that 
for generic $b$ (more precisely $b>2.303...$) their transition
temperatures are higher than $T_2$.

Since we already know $\mu>1$ (Eq \ref{mubigger1}),
we have to distinguish two cases

(i) if the tail exponent $\mu$ satisfies $1<\mu<2$,
then the second moment diverges at criticality $\overline{z^2}=+ \infty$
and we have the strict inequality $T_c<T_2$.

(ii) if the tail exponent $\mu$ satisfies $\mu>2$,
then the second moment is finite at criticality. The only possible
finite stable value is for the ratio $r_2$ is $r_2=b-1$.
 The critical temperature then
coincides with the transition temperature of the second moment $T_c=T_2$.

The scenario (i) is the most plausible, since 
the possibility (ii) would require some 'fine tuning' in some sense :
as explained in the introduction, the temperature only appears in the initial
condition (Eq \ref{zrecursioninitial}) of the renormalization ;
any initial temperature $T>T_c$ flows towards the high temperature fixed-point,
any initial temperature $T<T_c$ flows towards the low temperature fixed-point,
so that $T_c$ is defined as the only initial temperature from which the critical
distribution $Q_c(z)$ is accessible. The critical distribution $Q_c(z)$
has to satisfy the self-consistent equation of Eq \ref{selfmudp} 
to be stable. If (ii) were true, the distribution $Q_c(z)$
 should in addition satisfy a second completely independent
condition $r_2=b-1$, which seems unlikely.

In conclusion, we expect that the exponent $\mu$ introduced in 
 Eq. \ref{powerlawqcdp} satisfies
\begin{eqnarray}
 1<\mu<2
\label{rangeDPmu}
\end{eqnarray}
This is in agreement with the numerical simulations
presented below in section \ref{secdpnume}.

\subsubsection{Right tail behavior of the free-energy distribution}

Let us introduce the right-tail exponent $\eta_c$
\begin{eqnarray}
\ln P_c(f)  \opsimeq_{f \to +\infty} - \gamma' f^{\eta_c'} +... 
\label{righttailDP}
\end{eqnarray}

In the region where $f \to +\infty$, one has effectively
the high-temperature recursion 
\begin{eqnarray}
f \opsimeq \frac{ 1 }{b} \sum_{i=1}^{2b} f^{(i)}
\label{recursionDPcritiright}
\end{eqnarray}
where all free-energies are large.
A saddle-point analysis with the right tail of Eq. \ref{righttailDP}
shows that the only stable right exponent $\eta_c'$ should
satisfy $b = 2^{\eta_c'-1}$
i.e.
\begin{eqnarray}
\eta_c'=1+ \frac{\ln b}{\ln 2}
\label{etacright}
\end{eqnarray}

\subsection{ Critical exponent $\nu$ }

To compute the critical exponent $\nu$, we consider a small perturbation in the fluctuating
part of the free-energy of Eq \ref{fluctuatingfree}
\begin{eqnarray}
-\beta_c f_n^{(a)} \equiv -\beta_c (F_n^{(a)} - \overline{ F_n}  )
= -\beta \Delta_c u^{(a)} + \delta_n^{(a)}
\label{fluctuatingfreebis}
\end{eqnarray}
where $\delta_n^{(a)}$ represent the random perturbations of zero mean
\begin{eqnarray}
\overline{ \delta_n  } =0
\label{conditionzeromean}
\end{eqnarray}
Equivalently, these variables $\delta_n$ 
represent the perturbation at linear order
of the variables of Eq \ref{fluctuatingz}
\begin{eqnarray}
z_n^{(a)} \equiv e^{- \beta_c f_n^{(a)} } =  z_c^{(a)} + \delta_n^{(a)} 
\label{fluctuatingzbis}
\end{eqnarray}
The linearization of the recursion of Eq. \ref{zreducedrecursion} around the fixed point, 
yields
\begin{eqnarray}
\delta_{n+1} \simeq \sum_{i=1}^{2b}
 \frac{z_c^{(i)}}{ \cal B } \delta_{n}^{(i)}
\label{epsDP}
\end{eqnarray}
where $z_c^{(i)}$ are distributed with the critical distribution $Q_c(z)$.

As in the wetting case, the recurrence of Eq \ref{epsDP} 
 coincides with the recurrence 
describing a directed polymer on a Cayley tree
 \cite{Der_Spo,Der_review}, with the following differences

(i) the variables $\delta_n$ are random variables of zero mean 
(Eq \ref{conditionzeromean}), which is equivalent to 
 a  directed polymer model with random signs studied in
\cite{cayleycomplex}

(ii) more importantly, the random weights $\frac{z_c}{b}$
are distributed with the fixed point distribution $Q_c(z)$
presenting a broad power-law tail
in $1/z_c^{1+\mu}$ with $1<\mu<2$
(instead of $0<\mu<1$ for the wetting case).

Reasoning as in the wetting case, any 
 narrow symmetric distribution ${\cal P}_0(\delta_0)={\cal P}_0(-\delta_0)$
will lead to power-law tails of index $(1+\mu)$ after
one iteration. The study of the evolution of these tails
by iteration yields that the corresponding Lyapunov exponent $v$ 
will be determined by an equation similar to Eq. \ref{resvelocitywett}
\begin{eqnarray}
e^{\mu v}  =  \frac{2 b  \overline{ z_c^{\mu} } }{{\cal B}^{\mu}}
+ \frac{2 b }{{\cal B}^{\mu}} 
 \int_0^{+\infty} dy \frac{P_{\infty}(y)}{B_+} \ y^{\mu}
= 1+ 
\frac{2 b }{{\cal B}^{\mu}} 
 \int_0^{+\infty} dy \frac{P_{\infty}(y)}{B_+} \ y^{\mu} 
\label{eqvdp}
\end{eqnarray}
where we have used Eq. \ref{selfmudp}, 
in terms of the stationary distribution $P_{\infty}(y)$
of the rescaled process associated to Eq \ref{epsDP}
\begin{eqnarray}
y_{n+1} = e^{-v} \sum_{i=1}^{2b}  \frac{z_c^{(i)}}{{\cal B} } y_{n}^{(i)}
\label{epsdprescal}
\end{eqnarray}
The correlation length exponent then reads $\nu=\frac{\ln 2}{v}$.
As in the wetting case, the presence of the second term in Eq \ref{eqvdp}
is crucial to obtain a positive $v$ and a finite $\nu$.

\section{ Numerical study of the directed polymer transition }

\label{secdpnume}

As for the wetting transition (see Section \ref{wettnume}), 
we have used the 'pool-method' with a pool number $N=4.10^7$
to study the transition
of the hierarchical lattice of
 branching ratio $b=5$ with initial Gaussian energies (Eq. \ref{gaussian}).
The exact bounds on the critical temperature are  \cite{Coo_Der}
\begin{eqnarray}
T_0(b) =  \frac{1}{ \left[ 2 \ln b \right]^{\frac{1}{2}}}
\simeq 0.557...  \leq  T_c(b) \leq 
T_2(b) = \frac{1}{ \left[ \ln (b-1) \right]^{\frac{1}{2}}} \simeq 0.849..
\label{boundsgauss}
\end{eqnarray}
In \cite{Coo_Der}, the phase transition
has been studied numerically via the specific heat and the 
overlap. In this paper, we characterize the transition
via the statistics of free-energy and energy.
As in the wetting case, 
this allows to locate very precisely the pool-dependent critical temperature
and to measure the divergence of the correlation length $\xi(T)$
above and below $T_c$.

\subsection{ Flow of the free-energy width  }

\begin{figure}[htbp]
%\begin{figure}
\includegraphics[height=6cm]{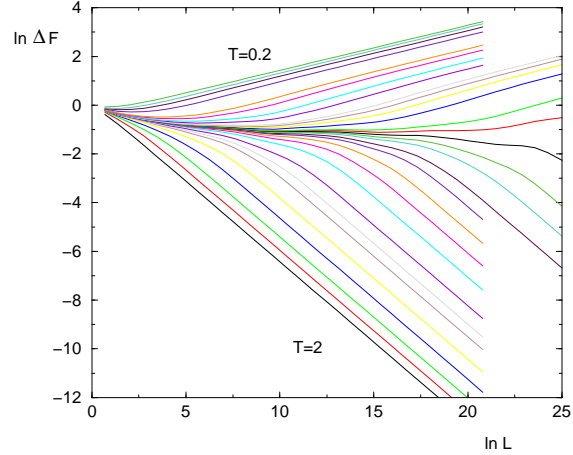}
\hspace{1cm}
\caption{(Color online) Directed polymer transition :
log-log plot of the width $\Delta F(L)$
of the free-energy distribution as a function of $L$, for many
temperatures. }
\label{figb5freewidth}
\end{figure}

The flow of the free-energy width $\Delta F_L$ as $L$ grows
is shown on Fig. \ref{figb5freewidth} for many temperatures.
One clearly sees the two attractive fixed points.
For $T>T_c$, the free-energy width decays asymptotically 
with the exponent $\omega_{\infty}(b)$ introduced in Eq. \ref{fwidthhighdp}
\begin{eqnarray}
\Delta F(L) \simeq \left( \frac{L}{\xi_F^+(T)} \right)^{- \omega_{\infty}(b)}
 \ \ { \rm with } \ \ \omega_{\infty}(b=5)=
 \frac{\ln (b/2) }{ 2 \ln 2} = 0.6609..
\label{fwidthabove}
\end{eqnarray}
where $\xi_F^+(T)$ is the corresponding correlation length
that diverges as $T \to T_c^+$.

For $T<T_c$, the free-energy width grows asymptotically 
with the exponent $\omega_0(b)$ of the ground-state energy distribution
\begin{eqnarray}
\Delta F(L) \simeq \left( \frac{L}{\xi_F^-(T)} \right)^{ \omega_0(b)}
\ \ { \rm with } \ \ \omega_0(b=5) \simeq 0.186...
\label{fwidthbelow}
\end{eqnarray}
where $\xi_F^-(T)$ is the corresponding correlation length
that diverges as $T \to T_c^-$.

For each given pool,
the flow of free-energy width allows a very precise
determination of the pool-dependent critical temperature,
for instance in the case considered 
$0.77662 < T_c^{pool} < 0.77666$ which is significantly below the upper
bound $T_2$ of Eq \ref{boundsgauss}.

\begin{figure}[htbp]
%\begin{figure}
\includegraphics[height=6cm]{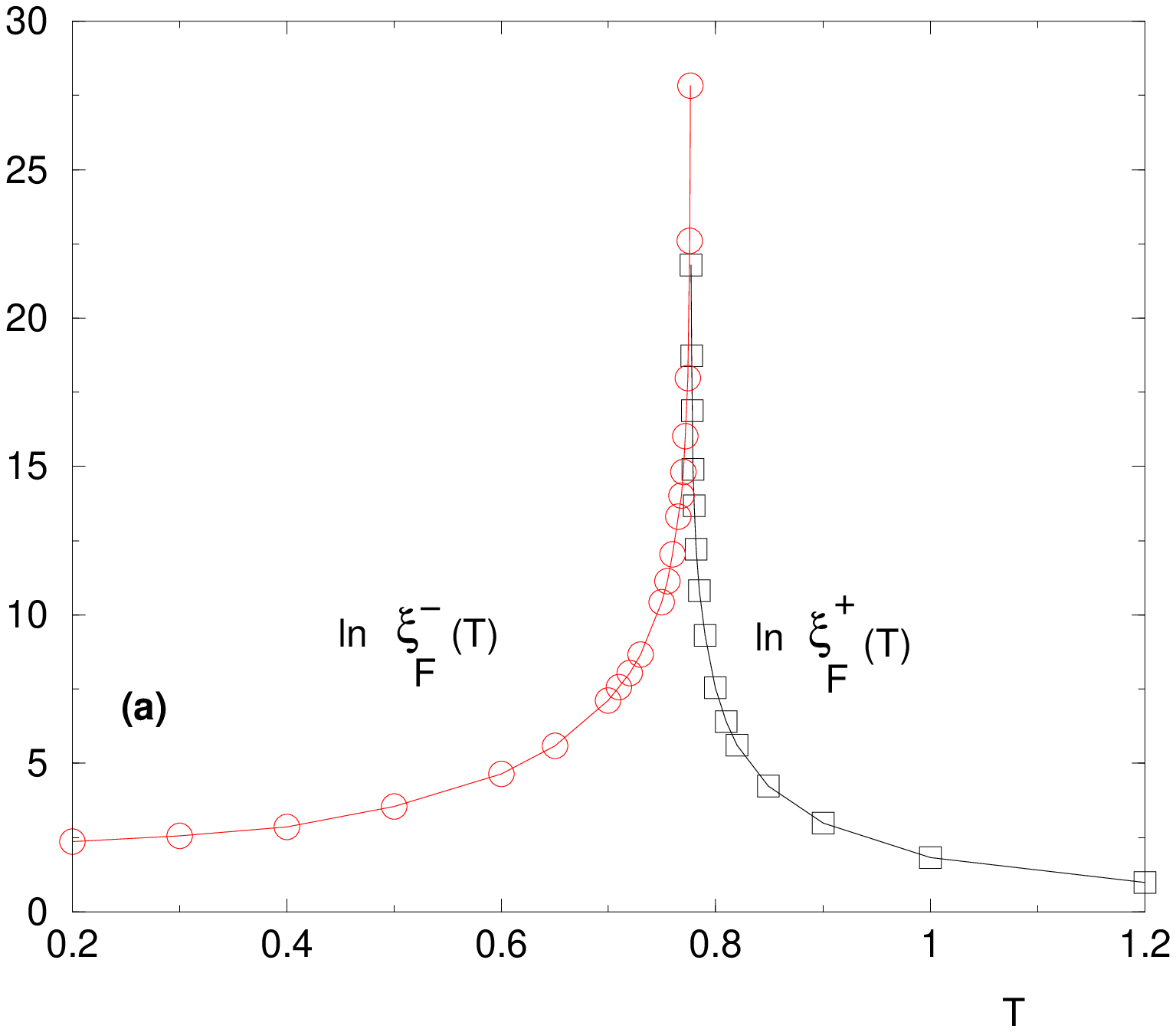}
\hspace{1cm}
\includegraphics[height=6cm]{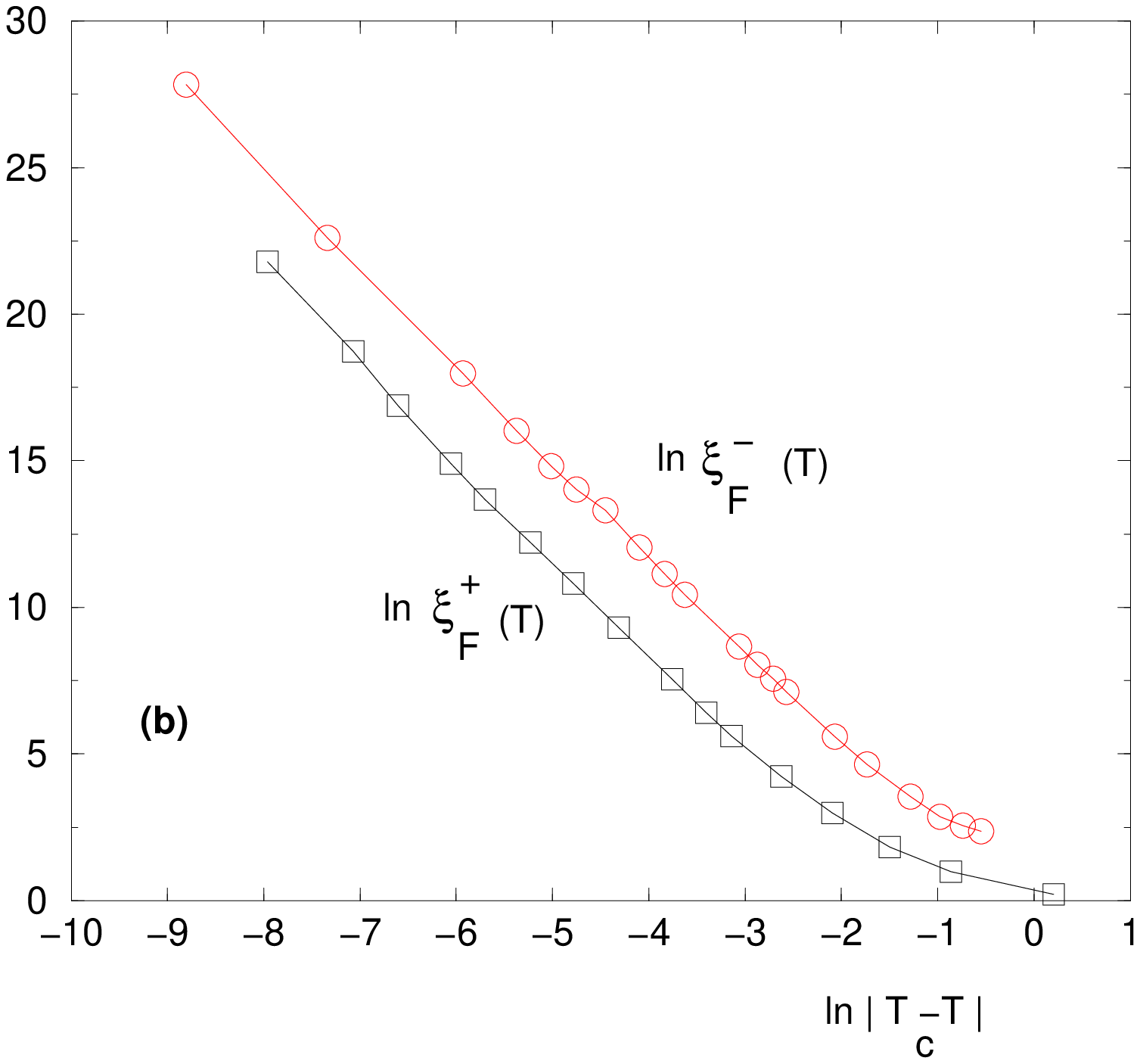}
\caption{(Color online)  Directed polymer transition :
Correlation length $\xi_F^{\pm}(T)$ as measured from the 
behavior of the free-energy width (Eqs \ref{fwidthabove} and
\ref{fwidthbelow}) (a) $\ln \xi_F^{\pm}(T)$ as a function of $T$
(b) $\ln \xi_F^{\pm}(T)$ as a function of $\ln \vert T_c-T \vert$ 
 : the asymptotic slopes are of order $\nu \sim 3.4 $
}
\label{figb5logxifree}
\end{figure}

\subsection{ Divergence of the correlation lengths $\xi_F^{\pm}(T)$ }

The correlation lengths $\xi_F^{\pm}(T)$ as measured from the free-energy
width asymptotic behaviors above and below $T_c$ (Eqs \ref{fwidthabove} and 
\ref{fwidthbelow} ) are shown on Fig. \ref{figb5logxifree} (a).
The plot in terms of the variable $\ln \vert T_c-T \vert$
shown on Fig. \ref{figb5logxifree} (b) indicate a power-law divergence
with the same exponent 
\begin{eqnarray}
\xi_F^{\pm}(T) \oppropto_{T \to T_c} \vert T-T_c \vert^{-\nu} 
\ \ { \rm with } \ \ \nu \simeq 3.4
\label{xifreenu}
\end{eqnarray}

\subsection{ Histogram of the free-energy } 

\begin{figure}[htbp]
%\begin{figure}
\includegraphics[height=6cm]{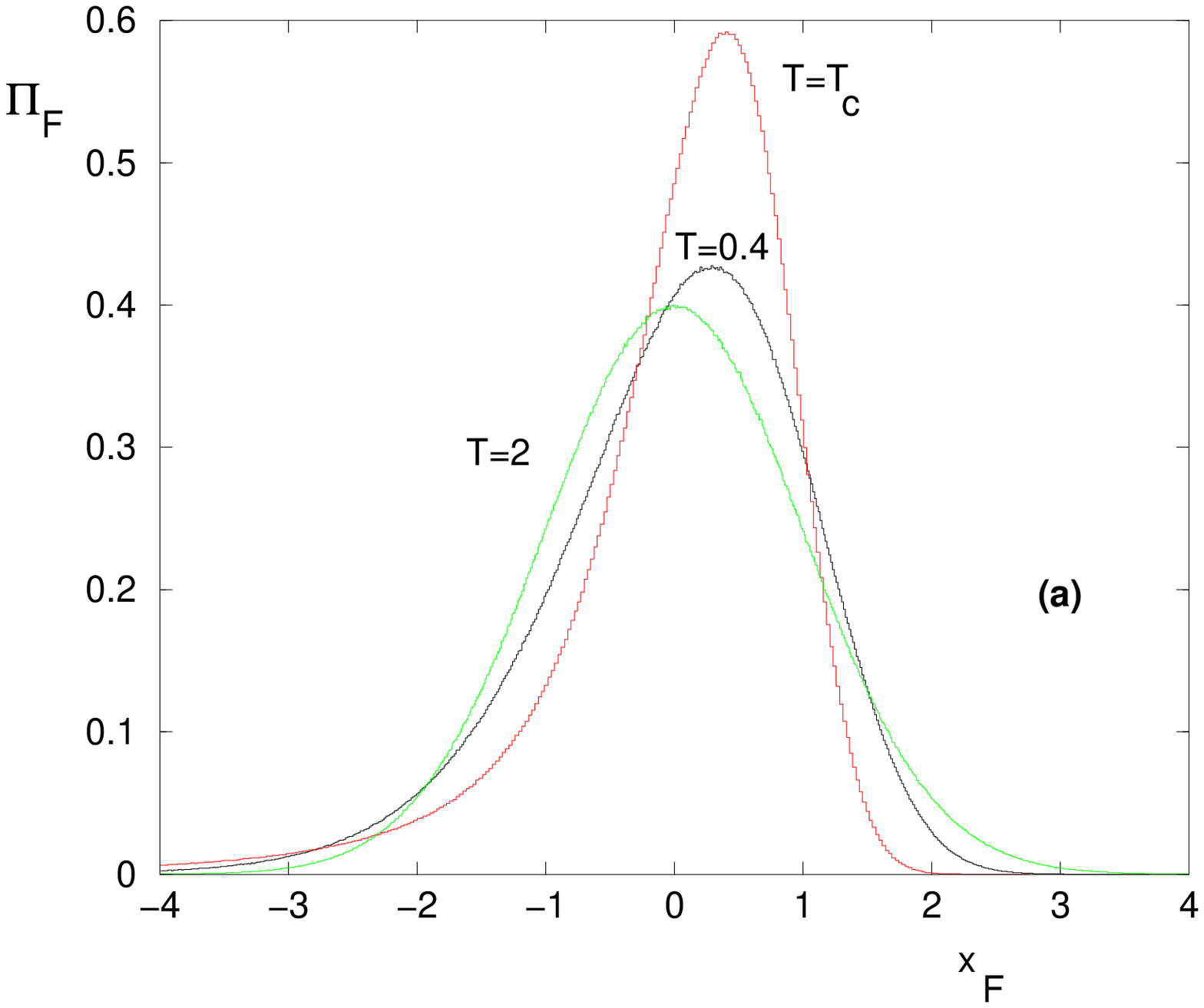}
\hspace{1cm}
\includegraphics[height=6cm]{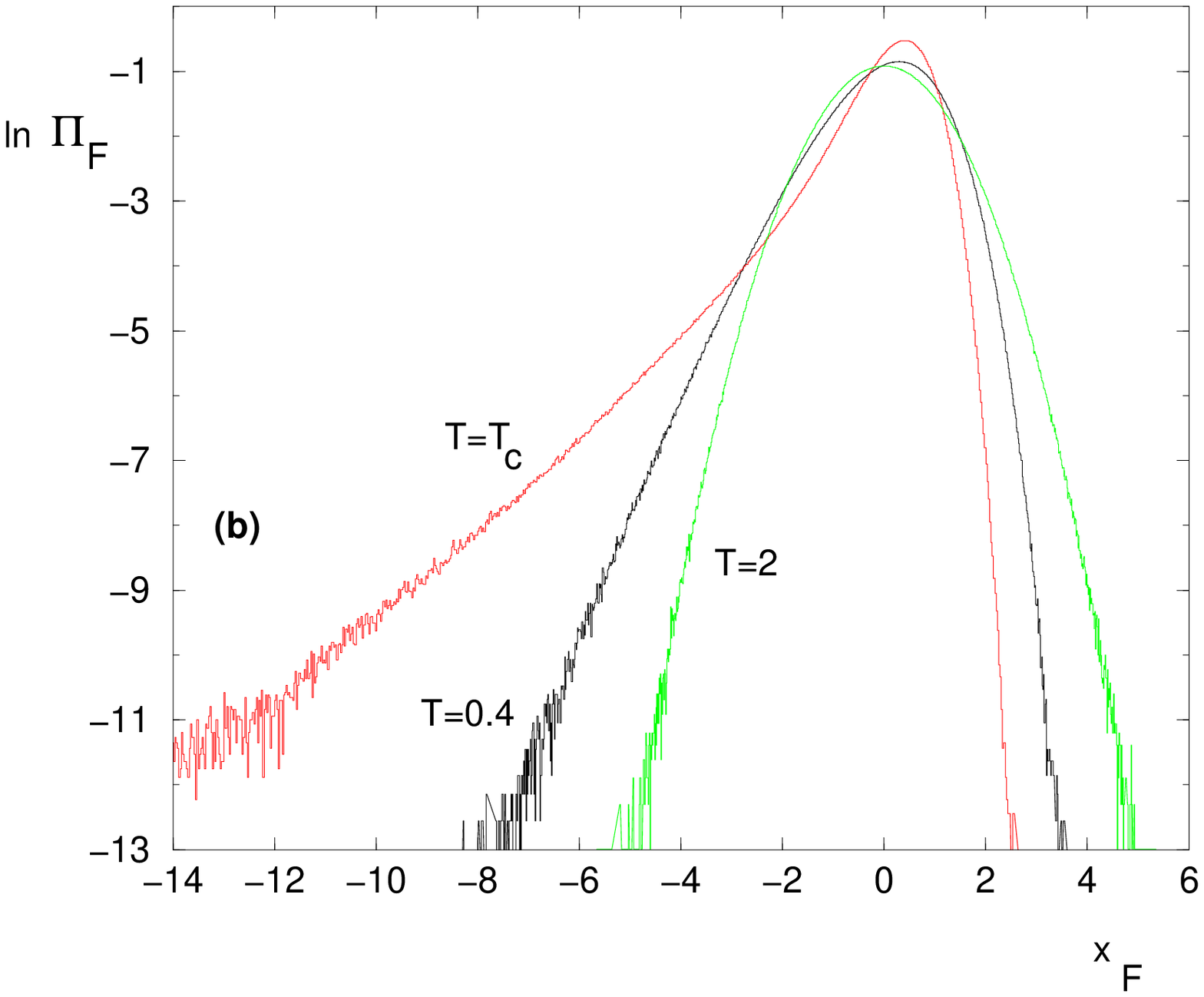}
\caption{(Color online)  Directed polymer transition :
Asymptotic distribution $\Pi_F$ of the rescaled free-energy 
$x_F= \frac{F-F_{av(L)}}{\Delta F(L)} $ 
in the low-temperature phase (here $T=0.4$), in the high-temperature phase
( here $T=2$) and at criticality (here $T_c^{pool}=0.77665$)
(a) Bulk representation 
(b) Log-representation to see the tails. }
\label{figb5freehisto}
\end{figure}

The asymptotic probability distribution $\Pi_F$ of the rescaled free-energy 
\begin{eqnarray}
x_F \equiv \frac{F-F_{av(L)}}{\Delta F(L)}
\end{eqnarray}
is shown on Fig \ref{figb5freehisto} for three temperatures :

(i) for $T>T_c$, it is a Gaussian in agreement with Eq \ref{uhigh}.

(ii) for $T<T_c$, it coincides with the ground state energy distribution.

(iii) at criticality, one clearly see that a left-tail 
develops with tail exponent $\eta_c =1$ in agreement with Eq \ref{etacdp}.
 The corresponding power-law exponent of Eq \ref{defimu}
of the fixed-point distribution of Eq. \ref{powerlawqc}
is of order 
\begin{eqnarray}
\mu \sim 1.6
\label{muvaluedp}
\end{eqnarray}
Again this measure is not precise as a consequence of the unknown
logarithmic correction in Eq. \ref{powerlawqc},
but it is in the expected interval of Eq \ref{rangeDPmu}.

\subsection{  Flow of the energy and entropy widths  } 

\begin{figure}[htbp]
%\begin{figure}
\includegraphics[height=6cm]{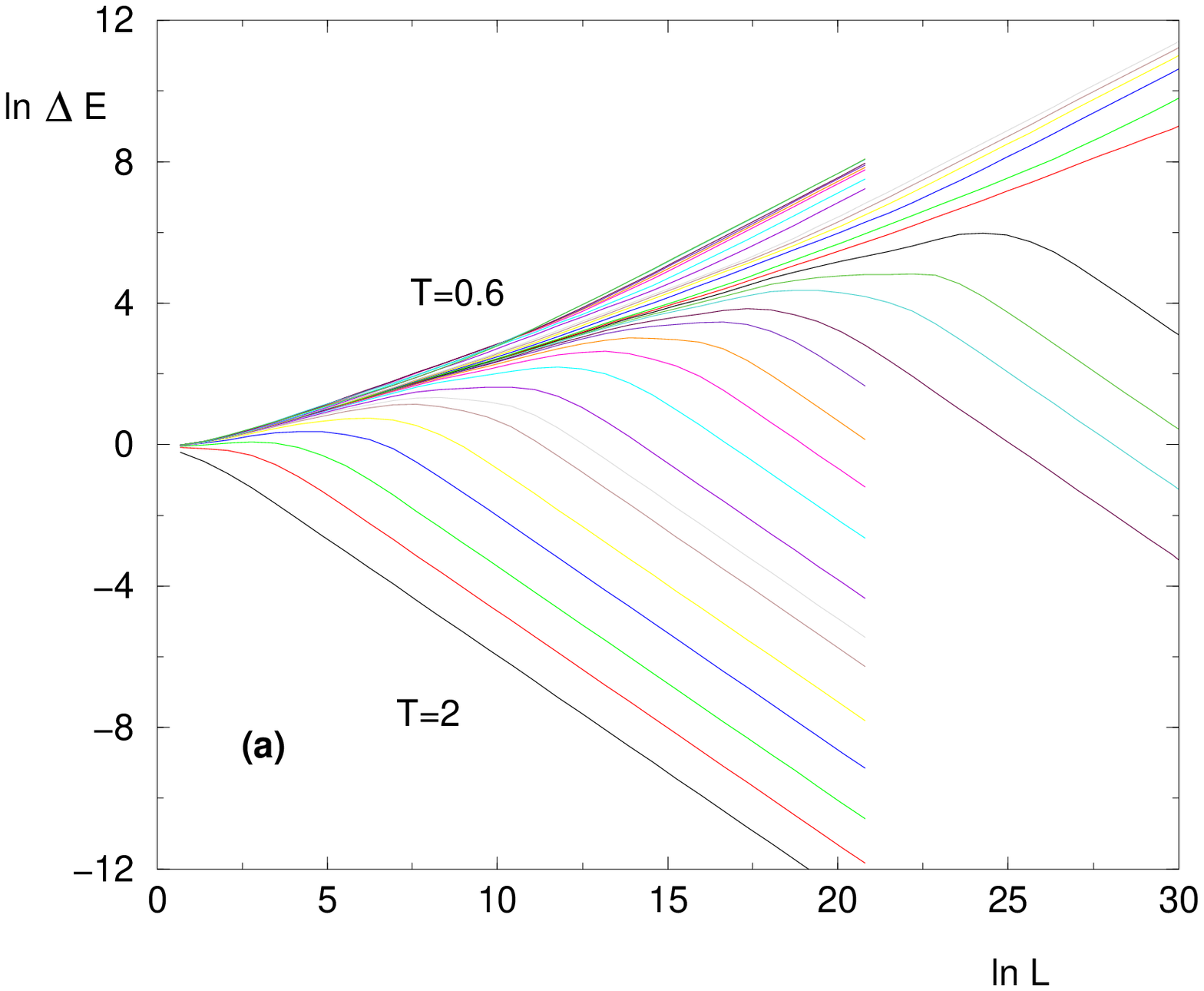}
\hspace{1cm}
\includegraphics[height=6cm]{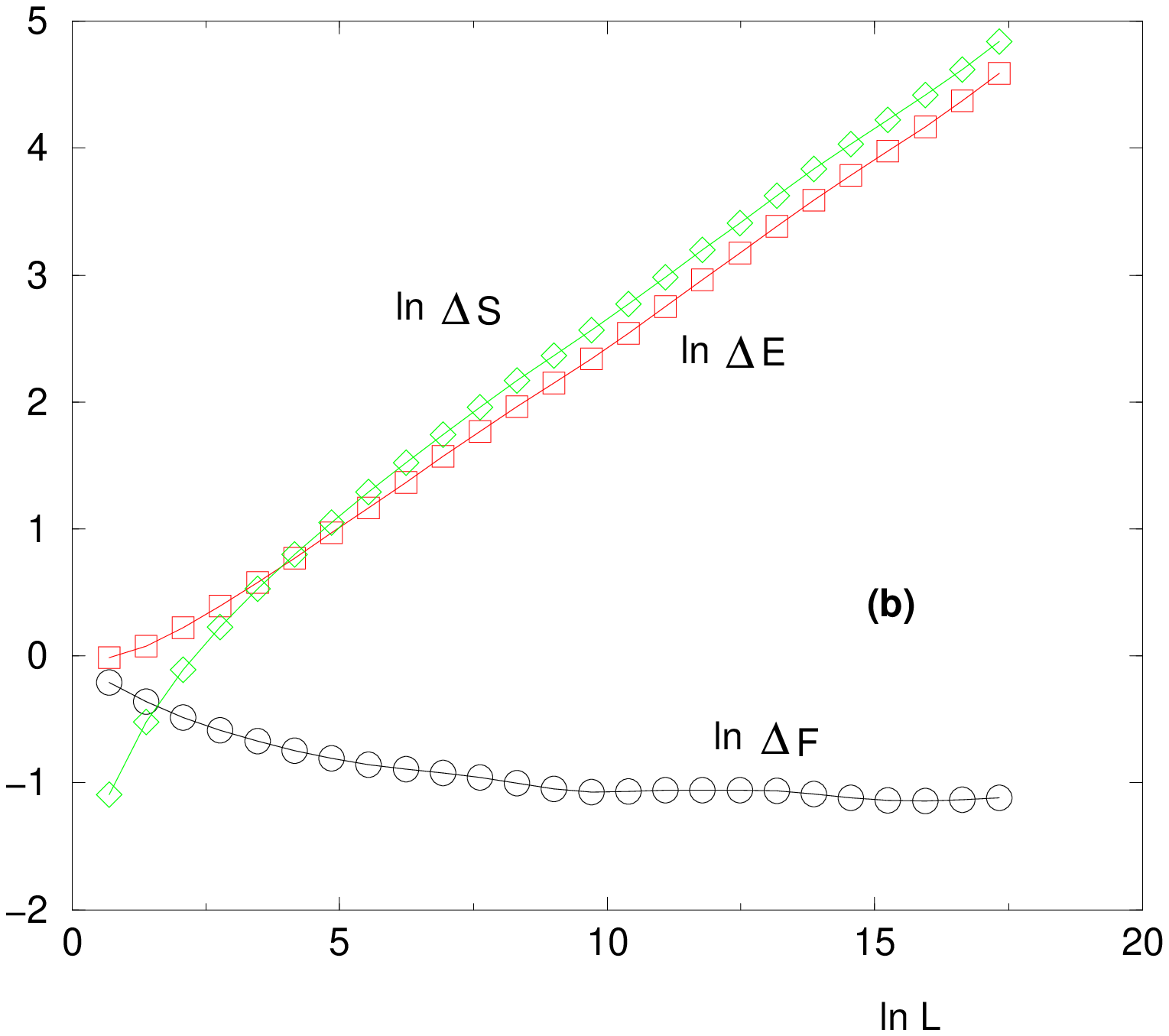}
\caption{(Color online)  Directed polymer transition :
Flow of the widths $\Delta E(L)$ 
of the energy distribution as $L$ grows
(a) $\ln \Delta E(L)$ as a function of $\ln L$ for many temperatures 
(b) Comparison
of $\ln \Delta E(L)$ , $ \ln \Delta S(L)$ and $ \ln \Delta F(L)$
as a function of $\ln L$ at criticality ($T_c^{pool}=0.77665$).}
\label{figb5enerwidth}
\end{figure}

The flow of the energy width $\Delta E(L)$ 
as $L$ grows is shown on Fig. \ref{figb5enerwidth} (a) for many temperatures
( the flow of the entropy width $\Delta S(L)$ is very similar at large scale).
For $T>T_c$, we find that these widths decay asymptotically 
with the same exponent $\omega_{\infty}(b)$ as the free-energy (Eq \ref{fwidthabove})
\begin{eqnarray}
\Delta E(L) \simeq L^{- \omega_{\infty}(b)} \\
\Delta S(L) \simeq L^{- \omega_{\infty}(b)}
\label{ewidthabove}
\end{eqnarray}
For $T<T_c$, in agreement with the Fisher-Huse droplet scaling theory 
for directed polymers \cite{Fis_Hus}, we find that 
these widths grow asymptotically 
with the exponent $1/2$
 which is bigger than the free-energy exponent $\omega_0(b)$ 
(Eq. \ref{fwidthbelow}) 
\begin{eqnarray}
\Delta E(L) \simeq L^{\frac{1}{2}} \\
\Delta S(L) \simeq L^{\frac{1}{2}}
\label{ewidthbelow}
\end{eqnarray}

Exactly at criticality, 
the free-energy $\Delta F(L)$ width converges towards a constant,
whereas the energy and entropy widths grow as power-laws 
(see Fig \ref{figb5enerwidth} b)
\begin{eqnarray}
\Delta E(L) \sim L^{y_c} \sim \Delta S(L) \ \ {\rm with } \ \ y_c \sim 0.29
\label{ewidthcriti}
\end{eqnarray}
This exponent is in agreement with the finite-size scaling relation $y_c=1/\nu$
with $\nu \sim 3.4$ (see Eqs \ref{xifreenu})

\subsection{ Divergence of the correlation lengths $\xi_E^{\pm}(T)$, 
$\xi_S^{\pm}(T)$ } 

\begin{figure}[htbp]
%\begin{figure}
\includegraphics[height=6cm]{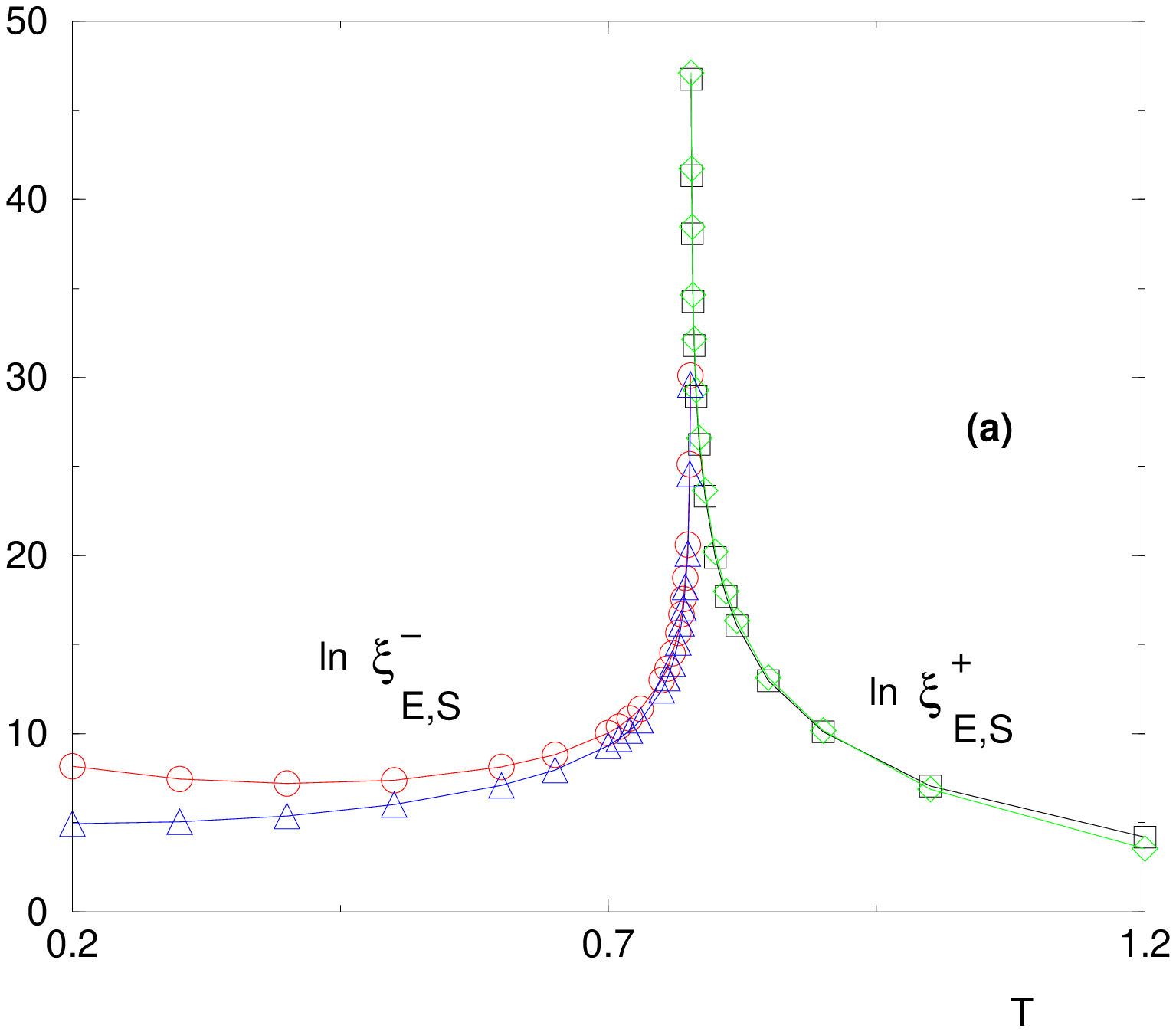}
\hspace{1cm}
\includegraphics[height=6cm]{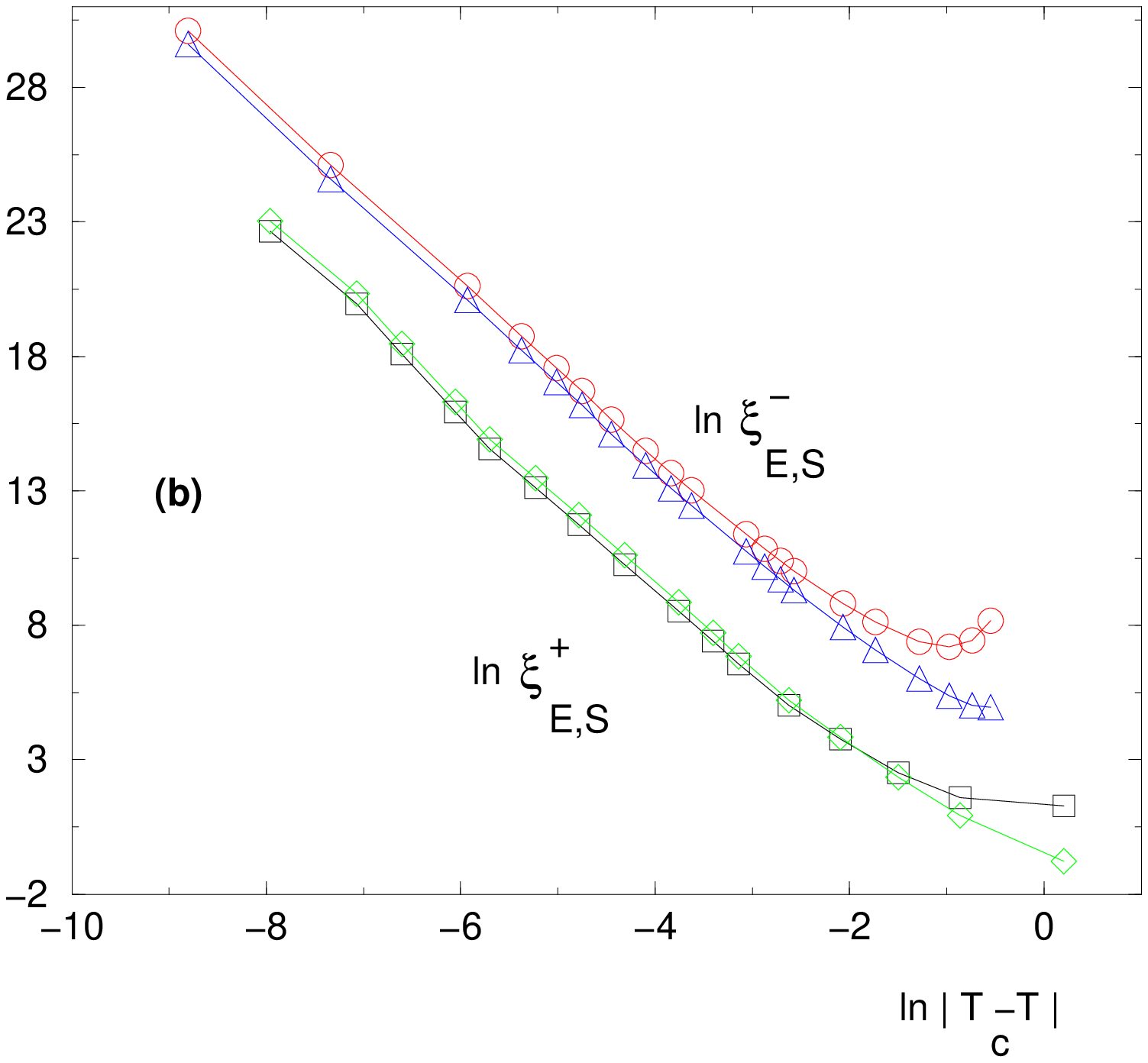}
\caption{(Color online)  Directed polymer transition :
Correlation length $\xi_E^{\pm}(T)$ 
(circles below and square above)
and $\xi_S^{\pm}(T)$ (triangles below and diamond above) as measured from the 
behavior of the energy 
and entropy widths 
(a) $\ln \xi_E(T)$ and $\ln \xi_S(T)$ as a function of $T$
(b) $\ln \xi_E(T)$ and $\ln \xi_S(T)$ as a function of $\ln \vert T_c-T \vert$:
the asymptotic slopes are of order $\nu \sim 3.4 $ as in
Fig. \ref{figb5logxifree} }
\label{figb5logxiener}
\end{figure}

According to the Fisher-Huse droplet 
scaling theory of spin-glasses \cite{Fis_Hus},
the singularities of the widths of energy and entropy as $T \to T_c^-$ 
is given by $(L/\xi(T))^{1/2}/(T_c-T)$.
We thus define the correlation lengths 
$\xi_E^+(T)$ and $\xi_S^+(T)$ by
\begin{eqnarray}
\Delta E(L) \simeq \frac{1}{T_c-T} \left( \frac{L}{\xi_E^-(T)} \right)^{\frac{1}{2}} \\
\Delta S(L) \simeq  \frac{1}{T_c-T} \left( \frac{L}{\xi_S^-(T)} \right)^{\frac{1}{2}}
\label{ewidthbelowcriti}
\end{eqnarray}

Similarly, for $T>T_c$, we define the corresponding correlation lengths 
$\xi_E^+(T)$ and $\xi_S^+(T)$ by the equations 
\begin{eqnarray}
\Delta E(L) \simeq \frac{1}{T-T_c}
\left( \frac{L}{\xi_E^+(T)} \right)^{- \omega_{\infty}(b)} \\
\Delta S(L) \simeq \frac{1}{T-T_c}
\left( \frac{L}{\xi_S^+(T)} \right)^{- \omega_{\infty}(b)}
\label{ewidthabovecriti}
\end{eqnarray}

The correlations lengths are shown on Fig. \ref{figb5logxiener} (a)
The plot in terms of the variable $\ln \vert T_c-T \vert$
shown on Fig. \ref{figb5logxiener} (b) indicate a power-law divergence
with the same exponent as in Eq. \ref{xifreenu}
\begin{eqnarray}
\xi_E^{\pm}(T) \oppropto_{T \to T_c} \vert T-T_c \vert^{-\nu} 
\ \ { \rm with } \ \ \nu \simeq 3.4
\label{xienernu}
\end{eqnarray}

\subsection{ Histogram of the energy } 

\begin{figure}[htbp]
%\begin{figure}
\includegraphics[height=6cm]{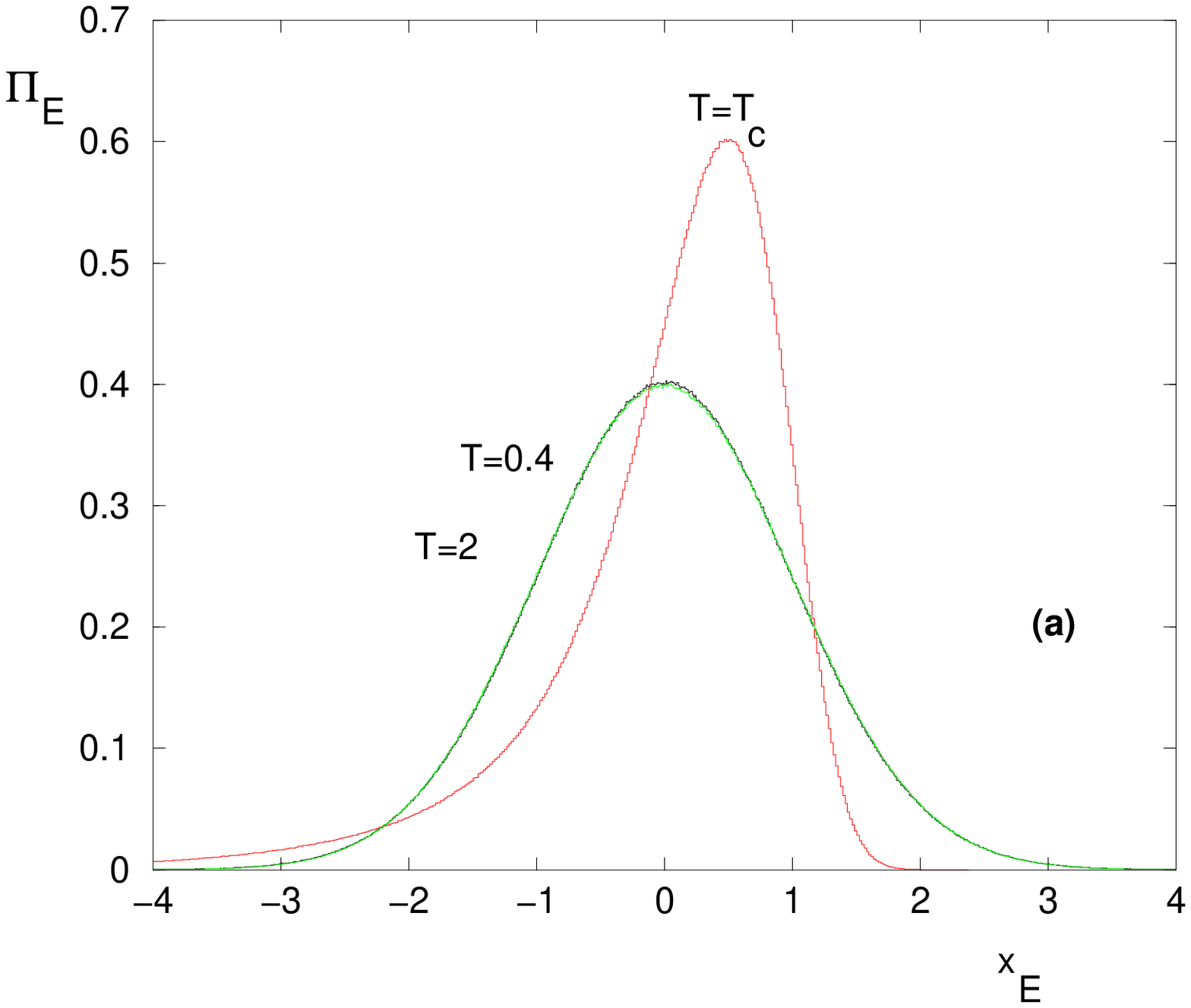}
\hspace{1cm}
\includegraphics[height=6cm]{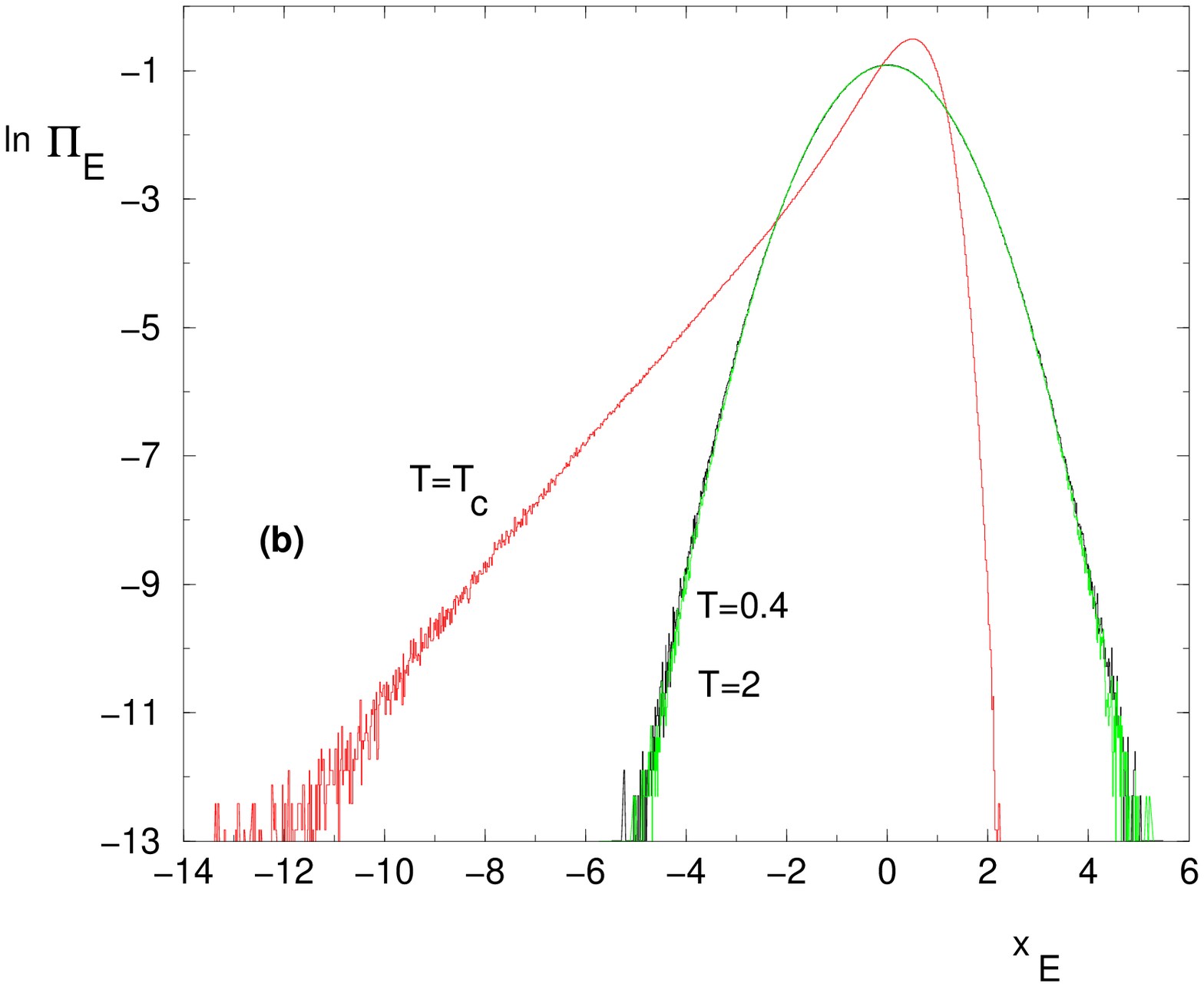}
\caption{(Color online)  Directed polymer transition :
Asymptotic distribution $\Pi_E$ of the rescaled energy 
$x= \frac{E-E_{av}}{\Delta E} $ 
in the low-temperature phase (here $T=0.4$), in the high-temperature phase
(here $T=2$) and at criticality ($T_c^{pool}=0.77665$)
(a) Bulk representation 
(b) Log-representation to see the tails }
\label{figb5enerhisto}
\end{figure}

The asymptotic probability distribution $\Pi_E$ of the rescaled energy 
\begin{eqnarray}
x_E \equiv \frac{E-E_{av(L)}}{\Delta E(L)}
\end{eqnarray}
is shown  for three temperatures on Fig \ref{figb5enerhisto}

(i) outside criticality, both for $T>T_c$ and $T<T_c$,
these distributions are Gaussian.

(ii) at criticality, the distribution is strongly non-gaussian
and asymmetric, with a left-tail of tail exponent $\eta_c =1$

\section{Comparison with corresponding results on hypercubic
lattices}

\label{comparison}

Since the exact renormalizations on the diamond lattice can also be considered
as approximate Migdal-Kadanoff renormalizations for hypercubic lattices,
it is interesting to discuss whether the results obtained for 
the wetting and the directed polymer on the diamond
lattice are qualitatively similar to the results for hypercubic lattices.

\subsection{ Similarities for $T<T_c$ }

The whole low-temperature phase of disordered systems
is usually characterized by the zero-temperature fixed point
where disorder fluctuations dominate.
For the disordered polymer models considered in this paper,
the free-energy fluctuations grow as power-law of the length
both for the diamond lattice and for hypercubic lattice
\begin{eqnarray}
\Delta F(L,T<T_c)  \oppropto_{L \to \infty}  L^{\omega_0} 
\label{zerotfixedpoint}
\end{eqnarray}
where $\omega_0$ is the exponent governing the fluctuations
of the ground state energy $E_0(L)$.
In the wetting case, this exponent has the simple value $\omega_0^{wett}=1/2$
that reflects the normal fluctuations of the $L$ random variables
defining the random adsorbing energies along the wall.
In the directed polymer case, the exponent $\omega_0$ is non-trivial
because the ground state configuration is the result of an optimization
problem.

\subsection{ Differences for $T>T_c$ }

The high temperature phase of disordered systems
is characterized by bounded disorder fluctuations,
but these fluctuations are not of the same order
on diamond lattices and on hypercubic lattices.
More precisely, for the disordered polymer models considered in this paper,
the free-energy fluctuations decays as a power-law
on the diamond lattices, whereas they
 remain of order $O(1)$ on hypercubic lattices
\begin{eqnarray}
{\rm Diamond : }
\ \ \ \ \Delta F(L,T>T_c) && \oppropto_{L \to \infty}  L^{-\omega_{\infty}(b)} \\ 
{\rm Hypercubic : }
\ \ \ \ \Delta F(L,T>T_c) && \oppropto_{L \to \infty}  O(1) 
\label{hightfixedpoint}
\end{eqnarray}
This difference seems to come from the 
boundary conditions :
(i) on the diamond lattice, the polymer is fixed at the two extreme points,
 but by the iterative construction of the lattice,
the coordinence of these two extreme points grow with the number $n$
of generations, so that it is possible to have a very efficient
averaging even near the boundaries
(ii) on the hypercubic lattices, the boundary conditions
are sufficient to produce free-energies fluctuations
of order $O(1)$ : the fixed origin has a finite coordinence,
and the fluctuations of order $O(1)$
of the random variables near this origin do not disappear as $L \to \infty$.

\subsection{ Differences at criticality }

On the hierarchical lattice, the free-energy fluctuations
of the disordered polymer considered here
are of order $O(1)$ at criticality 
\begin{eqnarray}
{\rm Diamond : }
\ \ \ \ \Delta F (L, T=T_c)  \oppropto_{L \to \infty} O(1)
\label{deltafdiamondcriti}
\end{eqnarray}
and it is the only possibility in the presence
of exact renormalizations :
if the free-energy width is growing, the flow will be
attracted at large scale
towards the zero-temperature fixed point of Eq. \ref{zerotfixedpoint},
 whereas if free-energy width is decaying, the flow will be
attracted towards the high-temperature fixed point of
 Eq. \ref{hightfixedpoint}. On hypercubic lattices, the free-energy fluctuations
$\Delta F(L,T_c) \sim L^{\omega_c}$ at criticality
are expected to be governed by a vanishing exponent
$\omega_c=0$, but they are not necessarily of order $O(1)$
because logarithms cannot be excluded, and have actually been found for
the directed polymer transition as we now explain.
Forrest and Tang \cite{Fo_Ta} have conjectured
from their numerical results on a growth model in the KPZ universality class
and from the exact solution of another growth model
that the fluctuations of the height of the interface
were logarithmic at criticality.
For the directed polymer model, this translates into
a logarithmic behavior of the free energy fluctuations at $T_c$
\begin{eqnarray}
{\rm Hypercubic   : }
\ \ \ \ \Delta F_{DP}^{3d} (L, T=T_c)  && 
 \oppropto_{L \to \infty}  (\ln L)^{\sigma} 
\label{deltafregularcriti}
\end{eqnarray}
where the exponent was measured to be $\sigma=1/2$ 
in $d=3$ \cite{Fo_Ta,Ki_Br_Mo,DP3d}
Further theoretical arguments
 in favour of this logarithmic behavior can be 
found in \cite{Ta_Na_Fo,Do_Ko}.
So the scaling of free-energy fluctuations at criticality
seem to be different
on hypercubic lattices and on diamond lattices.

Another related issue concerns the location of the critical temperature $T_c$
 with respect to upper bound $T_2$ :

(i) on the diamond lattice, 
the ratio $r_2$ of Eq \ref{defr2dp} is finite at $T_2$,
the ratio $z=Z/Z_{ann}$ is a finite random 
variable at $T_c$,
but the probability distribution of the corresponding partition
function presents a power law tail of index $(1+\mu)$
with $1<\mu<2$ (Eq. \ref{rangeDPmu}),
leading to the strict inequality $T_c<T_2$

(ii) on hypercubic lattices, the location of $T_c$ with respect
to $T_2$ is still controversial.
In \cite{DPdroplet}, we have argued that $T_c=T_2$ in dimension $d=3$,
because the divergence of $r_2 \sim e^{a \ln L} $ at $T_2$ 
is compatible with the logarithmic free-energy fluctuations
of Eq. \ref{deltafregularcriti}, provided the rescaled 
distribution of free-energy involves a left-tail exponent $\eta_c>1$,
as measured numerically in \cite{DP3d}. 
And in \cite{DP3dmultif}, we have found clear numerical evidence
from the statistics of inverse participation ratios
that the delocalization transition takes place at  
$T_2$.
However, other arguments are in favor of the strict inequality
$T_c<T_2$ in finite dimensions : 
a new upper bound $T^*<T_2$ was proposed in $1+3$ \cite{birkner},
and in \cite{camanes} the location of $T_c$
with respect to $T_2$ was shown to depend upon dimension and
probability distribution of the bond energies.
In particular for the gaussian distribution, the result $T_c<T_2$
is obtained only for $d \geq 5$ \cite{camanes}, but not for the case $d=3$
considered in numerical simulations \cite{DP3d,DP3dmultif}.

For the wetting transition in $1+1$ dimension, 
we are not aware of results concerning
the scale of free-energy fluctuations at criticality.

\bigskip

\bigskip

This comparison between the diamond lattice
and hypercubic lattice can be summarized as follows.
The free-energy fluctuations 
present analogous power-law behaviors in the low-temperature
phase (Eq. \ref{zerotfixedpoint}) but have different behaviors
in the high temperature phase (Eq \ref{hightfixedpoint}).
At criticality, the free-energy fluctuations could also
scale differently if logarithmic contributions
are present on regular lattices.

\section{Conclusion}

\label{conclusion}

In this paper, 
we have studied the wetting transition and the directed 
polymer delocalization transition on diamond hierarchical lattices.
These two phase transitions with frozen disorder 
correspond to the critical points
of quadratic renormalizations of the partition function.
We have first explained why the comparison with multiplicative
stochastic processes allows to understand the presence
of a power-law tail in the fixed point distribution
 $P_c(z) \sim \Phi(z)/z^{1+\mu}$ as $z \to +\infty$ 
( up to some sub-leading logarithmic
function $\Phi(z)$) so that all moments $z^{n}$ with $n>\mu$ diverge.
The exponent $\mu$ is in the range $0<\mu<1$
for the wetting transition ( the first moment diverges $\overline{z}=+\infty$ 
 and the critical temperature is strictly below
the annealed temperature $T_c<T_{ann}$) and is the range $1<\mu<2$
for the directed polymer transition ( the second moment 
diverges $\overline{z^2}=+\infty$
 and the critical temperature is strictly below
the transition temperature $T_2$ of the second moment.) 
We have then 
obtained that the linearized renormalization
around the critical point, which determines the exponent $\nu$,
coincides with the transfer matrix 
describing a directed polymer on the Cayley tree,
where the random weights determined by the fixed point
distribution $P_c(z)$ are broadly distributed.
We have shown that it induces important differences with respect
to the usual travelling wave solutions concerning more narrow distributions
of the weights \cite{Der_Spo,Der_review,cayleycomplex}, 
where the selected velocity only depends on the tail region.
Note that travelling waves also appear
in other renormalization approaches of random systems \cite{carpentier}.
Finally, we have presented detailed numerical results on the statistics 
of the free-energy and of the energy as a function of temperature
for the wetting
and the directed polymer transition on the diamond hierarchical lattice
with branching ratio $b=5$. In particular, we have shown that
the measure of the free-energy width $\Delta F(L)$
yields a very clear signature of the transition and allows to
measure the divergence of the correlation length $\xi^{\pm}(T)$
both below and above $T_c$ :
(i) for $T<T_c$, the free-energy width  
is governed by the zero-temperature exponent $\omega_0$
via $\Delta F(L) \sim (L/\xi_-(T))^{\omega_0}$;
(ii) for $T>T_c$, the free-energy width  
is governed by the high-temperature exponent $\omega_{\infty}$
via $\Delta F(L) \sim (L/\xi_+(T))^{-\omega_{\infty}}$.
From the point of view of histograms, the development of a left tail
with exponent $\eta_c=1$ at criticality is very clear and different
from histograms with exponent $\eta>1$ outside criticality.

\appendix

\section{ Reminder on multiplicative stochastic processes }

\label{multiplicative}

Multiplicative stochastic processes appears in many contexts,
in particular in one-dimensional disordered systems,
such as random walk in random potentials \cite{Kesten,Der_Pom,Bou}
or random spin chains \cite{Der_Hil,Cal}
In this Appendix, we recall some useful results concerning 
the following recurrence of random variables $X_n$
\begin{eqnarray}
X_{n+1}= a_n X_n + b_n
\label{msp}
\end{eqnarray}
where $(a_n, b_n)$ are positive independent random numbers.
The condition to have a stationary
probability distribution $P_{\infty}(X)$
is 
\begin{eqnarray}
\overline{ \ln a} <0
\label{stabcondition}
\end{eqnarray}
The most important property of $P_{\infty}(X)$
is that it presents a power-law tail 
\begin{eqnarray}
P_{\infty}(X) \opsimeq_{X \to +\infty} \frac{C}{X^{1+\mu}}
\label{powermu}
\end{eqnarray}
where the exponent $\mu>0$ is determined by the condition
 \cite{Kesten,Der_Pom,Bou,Der_Hil,Cal}
\begin{eqnarray}
\overline{ a^{\mu}} =1
\label{mspmu}
\end{eqnarray}

To understand where this condition comes from,
one needs to write that $P_{\infty}(X)$ is stable via the iteration
of Eq. \ref{msp}
\begin{eqnarray}
P_{\infty}(X) = \int da {\cal P}(a) \int db \psi(b) \int dY P_{\infty}(Y)
\delta \left( X- (aY+b) \right)
= \int da {\cal P}(a) \int db \psi(b) \frac{P_{\infty}( \frac{X-b}{a})}{a}
\label{eqmsp}
\end{eqnarray}
where ${\cal P}(a)$ and $\psi(b)$ are the probability distributions of $a_n$ and $b_n$ 
respectively.
The stability of the power-law tail of Eq \ref{powermu}
in the region $X \to +\infty$ yields at leading order
\begin{eqnarray}
\frac{C}{X^{1+\mu}}  \simeq \int da {\cal P}(a) \int db \psi(b)
 a^{\mu} \frac{C}{X^{1+\mu}} = \overline{ a^{\mu}} \ \ \frac{C}{X^{1+\mu}}  
\label{eqmspasymp}
\end{eqnarray}
yielding the condition of Eq. \ref{mspmu}

\end{document}